\documentclass[usenatbib,lscape]{emulateapj}
\usepackage{graphicx,ifthen,url,float,lscape,color,ulem}
\newcommand {\kms}{km s$^{-1}$}

\newcommand{\herschel}{\mbox{\it Herschel}}
\newcommand{\herschellong}{\mbox{\it Herschel Space Observatory}}
\newcommand{\spitzer}{\mbox{\it Spitzer}}

\newcommand {\lya}{Ly$\alpha$}
\def\ltsima{$\; \buildrel < \over \sim \;$}
\def\simlt{\lower.5ex\hbox{\ltsima}}
\def\gtsima{$\; \buildrel > \over \sim \;$}
\def\simgt{\lower.5ex\hbox{\gtsima}}
\newcommand {\uJy}{$\mu$Jy}
\newcommand {\um}{$\mu$m}

\def\um     {$\mu$m}
\def\ts     {\thinspace}
\def\kms    {\ifmmode{{\rm \ts km\ts s}^{-1}}\else{\ts km\ts s$^{-1}$}\fi}
\def\msol   {\ifmmode{{\rm M}_{\odot}}\else{M$_{\odot}$}\fi}
\def\lsol   {\ifmmode{{\rm L}_{\odot}}\else{L$_{\odot}$}\fi}
\def\zsol   {\ifmmode{{\rm Z}_{\odot}}\else{Z$_{\odot}$}\fi}
\def\etal   {{\rm et\ts al.}}
\def\ci     {\ifmmode{{\rm C}{\rm \small I}}\else{C\ts {\scriptsize I}}\fi}
\def\hi     {\ifmmode{{\rm H}{\rm \small I}}\else{H\ts {\scriptsize I}}\fi}
\def\hh     {\ifmmode{{\rm H}_2}\else{H$_2$}\fi}
\def\cone {\ifmmode{{\rm C}{\rm \small I}(^3\!P_1\!\to^3\!P_0)}
     \else{C\ts {\scriptsize I}{\small$(^3\!P_1\!\to\,^3\!P_0)$}}\fi}
\def\ctwo {\ifmmode{{\rm C}{\rm \small I}(^3\!P_2\!\to\,^3\!P_1)}
     \else{C\ts {\scriptsize I}{\small$(^3\!P_2\!\to\,^3\!P_1)$}}\fi}
\def\cij {\ifmmode{{\rm C}{\rm \small I}\,(^3P_i\to^3P_j)}\else{C\ts {\scriptsize I}\,{\small$(^3P_i\to^3P_j)$}}\fi}
\def\cii    {\ifmmode{{\rm C}{\rm \small II}}\else{C\ts {\scriptsize II}}\fi}
\def\tex {\ifmmode{{T}_{\rm ex}}\else{$T_{\rm ex}$}\fi}
\def\tmb {\ifmmode{{T}_{\rm mb}}\else{$T_{\rm mb}$}\fi}
\def\tkin {\ifmmode{{T}_{\rm kin}}\else{$T_{\rm kin}$}\fi}
\def\microns {\ifmmode{\mu{\rm m}}\else{$\mu$m}\fi}
\def\nhh   {\ifmmode{n({\rm H}_2)}\else{$n$(H$_2$)}\fi}

\newcommand{\iruv}{$L_{\rm IR}$/$L_{\rm UV}$}
\newcommand{\irx}{IRX}
\newcommand{\irxb}{IRX--$\beta$}
\newcommand{\iras}{{\it IRAS}}
\newcommand{\galex}{{\it GALEX}}

\newcommand{\msun}{{\rm\,M$_\odot$}}
\newcommand{\sfr}{{\rm\,M$_\odot$\,yr$^{-1}$}}
\newcommand{\lsun}{{\rm\,L$_\odot$}}

\newcommand{\ha}{{\rm\,H$\alpha$}}

\shorttitle{Are Dusty Galaxies... Blue?}
\shortauthors{C.~M. Casey et al.}
\begin{document}

\title{Are Dusty Galaxies Blue? Insights on UV Attenuation from Dust-Selected Galaxies}

\author{C.M.~Casey\altaffilmark{1}, N.Z.~Scoville\altaffilmark{2},
  D.B.~Sanders\altaffilmark{3}, N. Lee\altaffilmark{3},
  A. Cooray\altaffilmark{1}, S.L.~Finkelstein\altaffilmark{4},
  P. Capak\altaffilmark{5}, A. Conley\altaffilmark{6}, G. De
  Zotti\altaffilmark{7,8}, D. Farrah\altaffilmark{9},
  H. Fu\altaffilmark{10}, E. Le Floc'h\altaffilmark{11},
  O. Ilbert\altaffilmark{12}, R.J. Ivison\altaffilmark{13,14},
  T.T. Takeuchi\altaffilmark{15}}
\altaffiltext{1}{Department of Physics and Astronomy, University of California, Irvine, Irvine, CA 92697}
\altaffiltext{2}{California Institute of Technology, 1216 East California Boulevard, Pasadena, CA 91125}
\altaffiltext{3}{Institute for Astronomy, University of Hawai'i, 2680 Woodlawn Dr, Honolulu, HI 96822}
\altaffiltext{4}{Department of Astronomy, The University of Texas at Austin, Austin, TX 78712}
\altaffiltext{5}{Spitzer Science Center, California Institute of Technology, 1200 E California Boulevard, Pasadena, CA 91125}
\altaffiltext{6}{Center for Astrophysics and Space Astronomy 389-UCB, University of Colorado, Boulder, CO 80309}
\altaffiltext{7}{Osservatorio Astronomico di Padova, Vicolo dell'Osservatorio 2, 35122 Padova, Italy}
\altaffiltext{8}{SISSA, Via Bonomea 265, I-34136, Trieste, Italy}
\altaffiltext{9}{Department of Physics, Virginia Tech, Blacksburg, VA 24061}
\altaffiltext{10}{Department of Physics \&\ Astronomy, University of Iowa, Iowa City, IA 52242}
\altaffiltext{11}{CEA-Saclay, Orme des Merisiers, b\^{a}t. 709, F-91191 Gif-sur-Yvette Cedex, France}
\altaffiltext{12}{Aix Marseille Universit\'{e}, CNRS, Laboratoire d'Astrophysique de marseille, UMR 7326, 13388, Marseille, France}
\altaffiltext{13}{Institute for Astronomy, University of Edinburgh, Royal Observatory, Blackford Hill, Edinburgh, EH9 3HJ, UK}
\altaffiltext{14}{European Southern Observatory, Karl-Schwarzschild-Strasse 2, 85748 Garching bei M\"{u}nchen, Germany}
\altaffiltext{15}{Nagoya University, Division of Particle and Astrophysical Science, Furo-cho, Chikusa-ku, Nagoya 464-8602, Japan}

\label{firstpage}

\begin{abstract}
 Galaxies' rest-frame ultraviolet (UV) properties are often used to
directly infer the degree to which dust obscuration affects the
measurement of star formation rates. 
While much recent work has focused on calibrating dust attenuation in
galaxies selected at rest-frame ultraviolet wavelengths, locally and
at high-$z$, here we investigate attenuation in dusty, star-forming
galaxies (DSFGs) selected at far-infrared wavelengths.  By combining
multiwavelength coverage across 0.15--500\,\um\ in the COSMOS field, in
particular making use of \herschel\ imaging, and a rich dataset on
local galaxies, we find a empirical variation in the relationship
between rest-frame UV slope ($\beta$) and ratio of
infrared-to-ultraviolet emission (\iruv$\equiv$\irx) as a function of
infrared luminosity, or total star
formation rate, SFR.  Both locally and at high-$z$, galaxies above
SFR$\simgt$50\sfr\ deviate from the nominal \irxb\ relation towards
bluer colors by a factor proportional to their increasing IR
luminosity.
We also estimate contamination rates of DSFGs on high-$z$ dropout
searches of $\ll1$\%\ at $z\simlt4-10$, providing independent
verification that contamination from very dusty foreground galaxies is
low in LBG searches.
Overall, our results are consistent with the physical interpretation
that DSFGs, e.g. galaxies with $>50$\sfr, are dominated at all epochs
by short-lived, extreme burst events, producing many young O and B
stars that are primarily, yet not entirely, enshrouded in thick dust
cocoons. 
The blue rest-frame UV slopes of DSFGs are inconsistent with the
suggestion that most DSFGs at $z\sim2$ exhibit steady-state star
formation in secular disks.

\end{abstract}

\keywords{
galaxies: evolution $-$ galaxies: high-redshift $-$ galaxies: infrared
$-$ galaxies: starbursts $-$ submillimeter: galaxies}

\section{Introduction}

A key goal of extragalactic astrophysics is the accurate census and
measurement of the cosmic star formation rate density (SFRD) across
cosmic time.  The first measurements of the SFRD
\citep{tinsley80a,lilly95a,madau96a} revealed that the density of star formation
has decreased tenfold over the last seven billion years since
$z\sim1$. More recent measurements \citep*[like those summarized
  in][]{hopkins06a} show the nature of the SFRD at the highest
redshifts, reaching a plateau at $z\sim2$ and steadily declining at
earlier times \citep[now extended to $z\sim10$ with very deep
  near-infrared data, e.g.][]{oesch13a}.  However, there is one major
caveat to SFRD measurements derived from direct starlight in the
ultraviolet, optical and near-infrared, and that is the effect of dust
obscuration.

Detailed studies of dust attenuation in nearby star-forming regions
and star-forming galaxies
\citep[e.g.][]{calzetti94a,calzetti01a,meurer99a,overzier11a} have
long served as a calibration tool for understanding infrared
reprocessed emission in galaxies out to high-redshift.  One critical
tool has been the empirically observed tight correlation between
galaxies' rest-frame ultraviolet (UV) slope, defined as $\beta$, and
the ratio of infrared luminosity to UV luminosity at $\approx$1600\,\AA,
defined as \irx$\equiv$\,\iruv.  This local relationship between
$\beta$ and \irx\ has been widely used as a method to infer dust
obscuration, thus total star formation rates, in high-redshift
galaxies in the absence of far-infrared data
\citep[e.g.][]{bouwens09a}.  Thanks to substantial recent development
in deep, far-infrared instrumentation \citep*[see the review of ][for
  a complete discussion of far-infrared datasets used at
  high-redshift]{casey14a}, the robustness of this \irxb\ relationship
can now be tested at high-redshifts and high-luminosities.

Until recently, the analysis of the rest-frame UV properties of dusty
galaxies was limited to small, inhomogeneous samples.  Here we combine
the strengths of recent large area far-infrared surveys from
\herschel\ with the extensive 2\,deg$^2$ 30$+$ band
UV/optical/near-infrared photometry in the COSMOS field
\citep{capak07a,scoville07a,scoville13a} to investigate the rest-frame
UV characteristics of large samples of high-$z$ dusty star-forming
galaxies (DSFGs).  We compare the analysis of the $z>0$ DSFGs to
data of local galaxies of all luminosities
\citep{gil-de-paz07a,howell10a,u12a} to investigate the underlying
physical characteristics of dust attenuation at both low and
high-redshifts.  Section~\ref{sec:irxbhistory} describes some of the
relevant history of the \irxb\ relation, and
section~\ref{sec:selection} describes our galaxy samples$-$both nearby
and at $z>0$ in the COSMOS field$-$as well as presenting basic
calculations.  Our analysis of DSFGs in the \irxb\ plane is presented
in Section~\ref{sec:analysis}.  We address the possible contamination
of high-$z$ LBG dropout studies in Section~\ref{sec:highz}.  Our
discussion is later presented in Section~\ref{sec:discussion} and
conclusions in Section~\ref{sec:conclusions}.  Throughout, we assume a
Salpeter initial mass function \citep[][although we note conversion
  to a different IMF, thus different star formation rate, is
  straightforward]{salpeter55a} and a flat $\Lambda$CDM cosmology
\citep{hinshaw09a} with $H_{\rm 0}=71\,$km\,s$^{-1}$\,Mpc$^{-1}$ and
$\Omega_{\rm M}=0.27$.

\section{The \irxb\ Relation To-Date}\label{sec:irxbhistory}

The relationship between galaxies' relative dust attenuation, measured
as the ratio of IR (8--1000\,\um) to UV (1600\,\AA) luminosity
\iruv$\equiv$\irx, and their rest-frame ultraviolet color was first
studied in a sample of nearby starburst galaxies
\citep{meurer95a,meurer99a}.
Investigating the ultraviolet emission in nearby galaxies must be done from
space, and the first observations to contribute to this area came from
the {\it International Ultraviolet Explorer (IUE)} satellite
\citep{kinney93a}.  While the {\it IUE} played a critical role in
laying the groundwork for UV analyses of galaxies, and establishing
our understanding of the \irxb\ relationship, a key limitation of {\it
  IUE} data was its small aperture/field of view:
10\arcsec$\times$20\arcsec, typically much smaller than the full
spatial extent of nearby galaxies (a few arcminutes across).  It
should not then be surprising that the {\it IUE} data of
\citet{kinney93a} was focused on galaxy cores, which led to an
underprediction of galaxies' total UV luminosities.

In contrast to the UV, {\it IRAS} 12--100\,\um\ observations
\citep{neugebauer84a} and subsequent far-infrared observations of
nearby galaxies \citep[e.g.][]{dunne03a,kawada07a}, have been limited
to very large apertures.  Without matched apertures in the UV and IR,
the original analysis of the \irxb\ relation in \citet{meurer99a} was
biased by overestimating $IRX$, even though their selection of blue
compact dwarfs attempted to circumvent this problem.  Furthermore, the
{\it IUE} focus on only galaxy cores implied that there could also
likely be a UV color bias, with potential underestimation of the
global $\beta$ by only pointing towards the blue, UV-bright cores.
Thanks to later observations from the {\it Galaxy Evolution Explorer}
\citep[\galex;][]{morrissey07a} which provided wide field-of-view UV
photometry for the same nearby galaxies \citep[][]{gil-de-paz07a}, a
recent, revised analysis of the \citeauthor{meurer99a} relation finds
lower values of $IRX$ and redder UV slopes for the exact same galaxies
\citep[see][as well as some discussion in
  \citealt{overzier11a}]{takeuchi12a}.

Besides differences in photometry, it has long been known that
galaxies of different types present differently in the \irxb\ plane.
Young, metal-poor galaxies like the SMC and LMC are redder and less
dusty than starbursts, and normal galaxies lie between the young
SMC-type galaxies and compact blue starbursts
\citep{buat05a,buat10a,seibert05a,cortese06a,boissier07a,boquien09a,boquien12a,munoz-mateos09a,takeuchi10a,hao11a,overzier11a}.
Much of these differences are also likely caused by the differences in
interal attenuation curve, whether steep, shallow and with or without
the 2175\AA\ feature \citep{gordon00a,burgarella05a}.  \citet{kong04a}
provided a theoretical framework for the interpretation of these
differences by parametrizing galaxy types with a birthrate parameter,
$b$, defined as the ratio of present to past-averaged star formation
rate, whereby starbursts will have a much higher fraction of FUV
emission ($\sim$0.153\um) to NUV emission ($\sim$0.231\um) from O and
B stars.  While \citeauthor{kong04a} attribute the difference between
normal galaxies and starbursts to a difference in star-formation
histories, \citet{seibert05a} and \citet{cortese06a} argue using
\galex\ data that the differences are likely due to different dust
geometries.

Of particular interest for this paper is the observation that
ultraluminous infrared galaxies (ULIRGs) and related IR-bright galaxy
populations lie significantly {\it above} the \irxb\ relation, where
$IRX$ and $\beta$ have been claimed to be completely uncorrelated
\citep{goldader02a,burgarella05a,buat05a,howell10a,takeuchi10a}.  
\citet{howell10a} interpret sources lying above the \irxb\ relation as
having an excess of dust, and that the difference in IRX
($\Delta$\irx) from the expected relation represents the extent to
which the IR and UV emission is decoupled.  \citet{boquien09a} and
\citet{munoz-mateos09a} perform detailed studies of nearby galaxies
and conclude that both dust geometry and star-formation history have
substantial impact on the placement of galaxies on the \irxb\ relation.
Given the heightened importance of dusty ULIRGs to cosmic star
formation at $z\sim1$ and beyond
\citep{le-floch05a,caputi07a,casey12b,casey12c}, it seems crucial to
understand \irxb\ in dusty galaxies as well as unobscured galaxies.
To date, this has been unclear.

Beyond the nearby Universe, the \citet{meurer99a} \irxb\ relation has
played a fundamental role in estimating the amount of dust attenuation
at high-redshifts, particularly at $z>3$ where direct infrared
observations are unavailable
\citep{ouchi04a,stanway05a,hathi08a,bouwens09a}.  For example,
\citet{bouwens09a} use measurements of the rest-frame UV slope $\beta$
of very faint near-IR detected galaxies at $6<z_{\rm phot}<10$ to
constrain the dustiness of the early Universe.  Because
\citeauthor{bouwens09a} find that sources in high-$z$ surveys are
significantly bluer than low-$z$ galaxies, they conclude that the dust
obscuration plays an insignificant role in galaxy evolution at $z>5$.
Given that the nominal \irxb\ relation is actually quite uncertain
(and based on potentially biased galaxy samples) this constraint on
high-$z$ dust obscuration needs independent verification from
infrared/submillimeter surveys.

Constraining dust obscuration (and star formation rates) at the
highest redshifts requires a thorough understanding and calibration of
the \irxb\ relation beyond the local Universe.  Unfortunately,
calibrations of the \irxb\ relation have been very limited due to
longstanding limitations and sensitivity of far-infrared
observations. %, only recently remedied \citep*[see recent review of ][
%  on dusty star-forming galaxies, DSFGs at high-$z$]{casey14a}.  The
The most thorough analysis of the \irxb\ relation at high-$z$ has come
from studies of spectroscopically confirmed $z\sim2$ Lyman Break
Galaxies \citep[LBGs;][]{reddy06a,reddy09a,reddy10a,reddy12a}.  While
earlier works (those before about $\sim$2010) largely relied on
indirect measurements of galaxies' far-infrared luminosity,
\citet{reddy12a} use some of the deepest pointings of the
\herschellong\ in the GOODS fields \citep{elbaz11a} to investigate the
direct far-infrared emission in LBGs.  Since very few LBGs are
directly detected with \herschel, they used a stacking analysis to
measure the characteristic $L_{\rm IR}$ for $z\sim2$ LBGs and found
$\langle L_{\rm IR}({\rm LBG})\rangle \sim$10$^{11}$\lsun; this
characteristic luminosity indicates that $\sim$80\%\ of the star
formation is obscured in L$_\ast$ galaxies at $z\sim2$.  This lines up
with expectation from the nominal \citet{meurer99a} \irxb\ attenuation
curve (in spite of its known problems caused by {\it IUE} aperture
limitations and lack of accommodation for `normal' type galaxies).
Recently, work on $z\sim4$ LBGs \citep{to14a}, using radio continuum
measurements instead of direct far-infrared data, show further
agreement with the \citeauthor{meurer99a} relation.  Similar to the
\citeauthor{reddy12a} results, \citet{heinis13a} explore the dust
attenuation law in large samples of UV-selected galaxies at $z\sim1.5$
and find a roughly constant \irx\ ratio over a wide range of UV
luminosity explored and an \irxb\ relation in line with local `normal'
star-forming galaxies.  Further stacking results beyond $z>3$
potentially hint at a breakdown in the relation for highly luminous
LBGs \citep{lee12a,coppin14a}.

While the \citeauthor{reddy12a} and \citeauthor{heinis13a} direct
far-infrared studies have shed valuable light on dust attenuation
calibrations at high-$z$, they both address the problem using
UV-selected galaxy samples, which preferentially might have bluer UV
colors and lower \irx\ values than the average galaxy at
high-redshift.  In contrast, \citet{penner12a} present an analysis of
the \irxb\ relationship for 24\um-selected dust-obscured galaxies
\citep[DOGs;][]{dey08a}, which have direct detections in the
far-infrared from \herschel-{\sc pacs} \citep{poglitsch10a}.  They
find, perhaps not surprisingly, that dustier galaxies have higher
\iruv\ ratios than those selected at UV wavelengths, even at matched
rest-frame UV slopes.  This is quite similar to earlier work on local
ULIRGs \citep{goldader02a,howell10a} which found similarly high
\iruv\ ratios, above expectation from \irxb.  This difference between
`normal' star-forming galaxies and infrared-luminous starbursts is
attributed to emergent UV emission not corresponding to the same
spatial regions of dust absorption and re-emission
\citep{trentham99a,papovich06a,bauer11a}.  Indeed, some select studies
describing the rest-frame UV and optical properties of submillimeter
galaxies \citep[SMGs, selected at 850\,\um;][]{smail97a} at
high-redshift reinforce this result by finding little correspondence
between UV luminosity, UV color, and FIR luminosity
\citep{frayer00a,smail04a,chapman05a,casey09b,walter12a,fu12a}.
This evidence has so far indicated that the nominal
\irxb\ relation should not be used in dusty galaxies (and vice versa,
that any relationship determined for dusty galaxies should not be
assumed for a general galaxy population), but what more can we learn?

%
%Also reference \citep{laird05a}

\section{Galaxy Samples and Calculations}\label{sec:selection}

We use two sets of galaxies in this analysis; the first is a set of
nearby galaxies ($z<0.085$) spanning the characteristic range of
galaxy environments in the local volume, with $L_{\rm
  bol}\sim$10$^{(8-12.5)}$\,\lsun\ and star formation rates
$\approx$0.01--100\sfr.  The second galaxy population consists of
far-infrared selected star-forming galaxies spanning a redshift range
$0<z<5$, with most at $z<3.5$.  The selection, data-sets, and
description of these samples follow.  The calculation of relevant
quantities like $L_{\rm UV}$, $L_{\rm IR}$, \irx\ and $\beta$ are
performed in a consistent manner between local and $z>0$
samples.

\subsection{Nearby Galaxies}

Our local dataset combines the 1034 nearby galaxies observed by
\galex\ included in \citet{gil-de-paz07a}, originally selected from
the \galex\ Nearby Galaxies Survey (NGS), and the 202 nearby GOALS
\citep{armus09a} LIRGs and ULIRGs observed by
\galex\ \citep{howell10a}, originally selected from the \iras\ Revised
Bright Galaxy Sample \citep[RBGS;][]{sanders03a}.  Note that the
\galex\ photometry for several of the sources in the \citet{howell10a}
are slightly revised in \citet{u12a} to account for highly irregular
morphologies associated with galaxy interactions.  The IR and UV
apertures are matched and taken as total.

It is worth noting that all of the original blue compact galaxies
used to derive the \irxb\ relation in \citet{meurer99a} are included
in the \citet{gil-de-paz07a} sample; most of them are significantly
redder (higher $\beta$ value) when measured with \galex, consistent with
other works that have since updated the nominal \irxb\ relation
\citep{takeuchi10a,takeuchi12a}.

\subsection{$0<z<3.5$ Dusty Galaxies}

We use a sample of $>$4000 \herschel-selected DSFGs in the COSMOS
field \citep{scoville07a,scoville13a,capak07a} as our $z>0$ sample,
which have full UV-to-radio photometric coverage.
\herschel\ \citep{pilbratt10a} coverage of COSMOS was carried out with
the {\sc pacs} \citep{poglitsch10a} and {\sc spire} \citep{griffin10a}
instruments as part of the PEP \citep{lutz11a} and HerMES
\citep{oliver12a} surveys.  The DSFGs' characteristics and selection
($>$3$\sigma$ in two or more of the five \herschel\ bands from
100--500\,\um) are described in detail in \citet{lee13a}, who present
their empirically determined spectral energy distributions from the UV
through the far-infrared.  The method we use to estimate galaxies'
deboosted far-infrared flux densities is described in
\citet{roseboom10a} with updates in \citet{roseboom12a}.
The sample can largely be regarded as a
luminosity-limited sample, as selection is unbiased with respect to
far-infrared SED shape (i.e. dust temperature); deboosted flux
densities are estimated to be complete above $S_{\rm
  deboosted}\approx$10\,mJy at 160\,\um, 250\,\um, 350\,\um, and 500\,\um\ and
above $\approx$6\,mJy at 100\,\um.  We refer the reader to
\citet{lee13a} for details on sample selection.  DSFGs' astrometry and
multiwavelength counterparts are identified via a cross-identification
method using 24\um\ and 1.4\,GHz position priors down to
$S_{24}=80$\,\uJy\ and $S_{\rm
  1.4GHz}=$65\,\uJy\ \citep{roseboom10a,roseboom12a}.  This technique
is predicated on the notion that high-redshift DSFGs should either be
24\um\ or 1.4\,GHz {\it detected}; this is an accurate assumption at
$z\simlt2$ \citep[with only a 3\%\ failure rate, see][]{magdis11a},
but the sample is largely largely incomplete at higher redshifts where
24\um\ and 1.4\,GHz surveys are relatively insensitive to detecting
$z>2$ galaxies.

Of the original 4546 {\it Herschel}-identified galaxies with 24\um\ or
1.4\,GHz counterparts, only 4165 have reliable photometric redshifts
\citep{ilbert09a}: $\approx$92\%\ of the total sample.  The accuracy
of the photometric redshifts in COSMOS is reported as $\sigma_{\Delta
  z/(1+z)}=0.012$.  It should be noted that the DSFGs without reliable
redshifts {\it do} exhibit different optical/near-IR photometry than
those with reliable redshifts.  Most are undetected in all but a few
bands.  Without redshifts, it is difficult to characterize this subset
of the population with respect to \irxb, beyond claiming that they
likely have \irx$>$3.

\begin{figure}
\centering
\includegraphics[width=0.99\columnwidth]{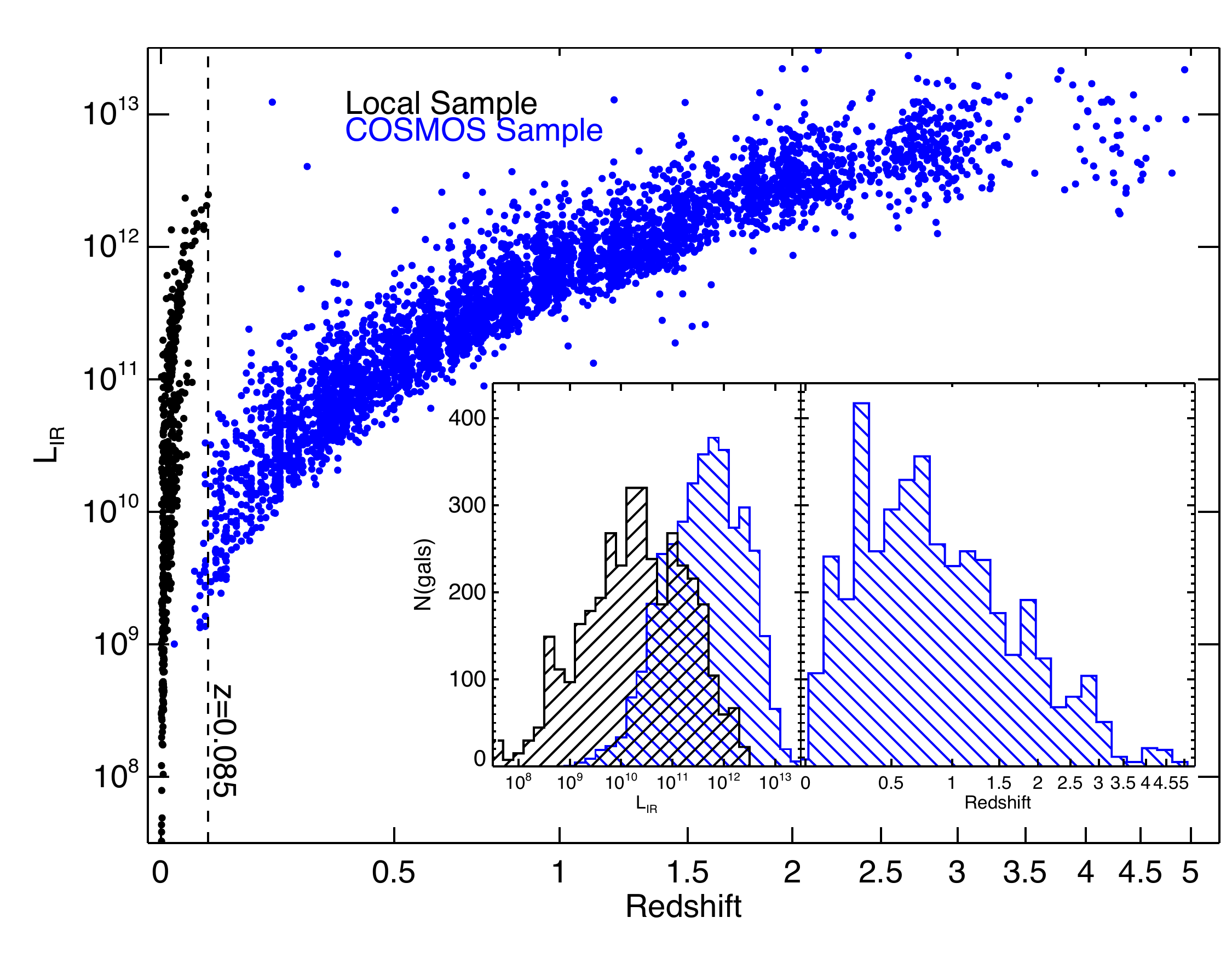}
\caption{IR luminosity against redshift for both the local sample
  (black points) and the COSMOS $z>0$ sample (blue points). Inset are
  histograms contrasting their IR luminosities (left inset) and
  redshifts (right inset).  The local sample sits below $z=0.085$
  while the median redshift for the COSMOS sample is 0.83, with
  60\%\ of the sample sitting at $0.1<z<1$ and 98\%\ below $z$ of
  3.5.}
\label{fig:nz}
\end{figure}

Since emission from active galactic nuclei (AGN) has the potential to
affect the results of this study$-$by potentially skewing the
rest-frame UV colors bluer and boosting the IR luminosities in the
mid-IR regime$-$we must remove sources with X-ray detections, strong
powerlaws in \spitzer\ IRAC and MIPS bands and sources with
unexpectedly high radio-to-FIR ratios.  Of the 4218 \herschel-detected
sources in COSMOS, only 5 are directly detected by XMM
\citep{brusa10a} and a further 95 are detected by {\it Chandra}
\citep{civano12a} suggesting a powerful AGN.  They are removed from
our analysis.  While obscured AGN are less likely to contaminate the
rest-frame UV colors of their host galaxies, we also remove 332
additional galaxies suspected of hosting such AGN, evidenced by a
strong powerlaw through \spitzer\ IRAC and MIPS bands
\citep{donley12a}.  All COSMOS DSFGs which exhibit very high
radio-to-FIR ratios have been identified as AGN through X-ray
selection or mid-infrared powerlaw.  Of the remaining 4165 DSFGs in
our study with photometric redshifts, a total of 432 ($\approx$10\%)
have been removed as potential AGN contaminants, leaving 3733 DSFGs
for our analysis.  Figure~\ref{fig:nz} plots the IR luminosities and
redshifts of the COSMOS DSFG sample alongside the local sample for
comparison.  While 98\%\ of the DSFG sample sits at $z<3.5$, the
median redshift for the COSMOS sample is 0.83, in close agreement with
the measured peak redshift for {\it Herschel}-bright galaxies
\citep{casey12b}.

\subsection{Calculations}\label{sec:calculations}

We fit $L_{\rm IR}$ consistently between the local and $z>0$
samples, from 8 to 1000\,\um\ and using all available infrared to
millimeter data at rest-frame 8--2000\,\um.  For the local sample,
this is primarily based on the four {\it IRAS} bands at 12\um, 25\um,
60\,\um\ and 100\,\um, with some additional data such as {\sc Scuba}
\citep{dunne03a} for the brightest nearby (U)LIRGs. The full detailed
photometry is described in \citet{u12a}.  The COSMOS sample includes
the five-band \herschel\ 100--500\,\um\ data,
24--70\,\um\ \spitzer\ data, and when available AzTEC 1.1\,mm data
\citep{scott08a,aretxaga11a} as well as {\sc Scuba-2} 450\,\um\ and
850\,\um\ data \citep{casey13a,roseboom13a}.  We use the far-infrared
SED fitting technique described by \citet{casey12a}; this has been shown
to produce $L_{\rm IR}$, $T_{\rm dust}$ and $M_{\rm dust}$ estimates
that are more accurate (when compared to direct interpolation of data)
yet fully consistent with template fitting methods popular in the
literature \citep*[e.g.][]{chary01a,dale02a}. 

\begin{figure*}
\centering
\includegraphics[width=0.99\columnwidth]{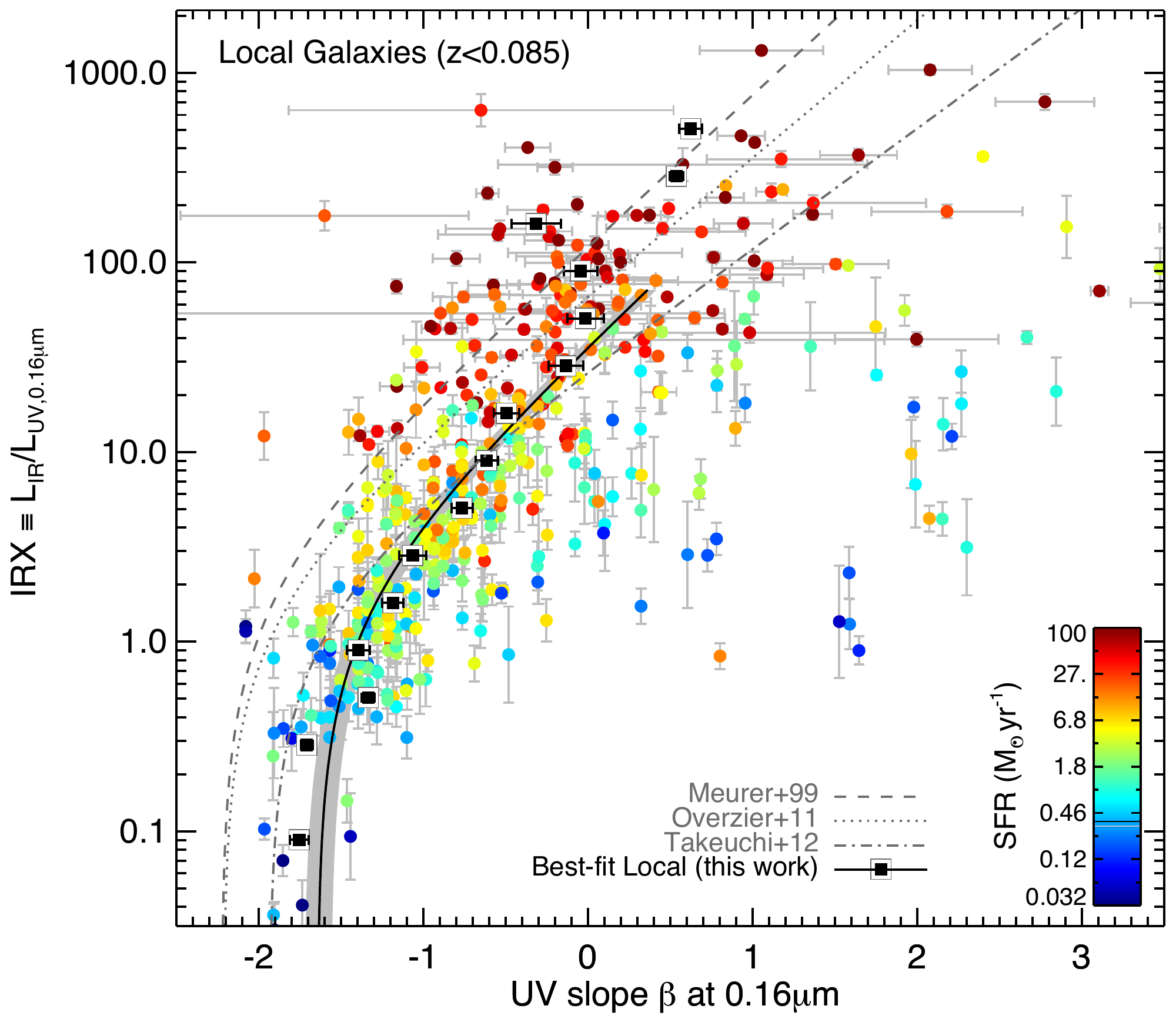}
\includegraphics[width=0.99\columnwidth]{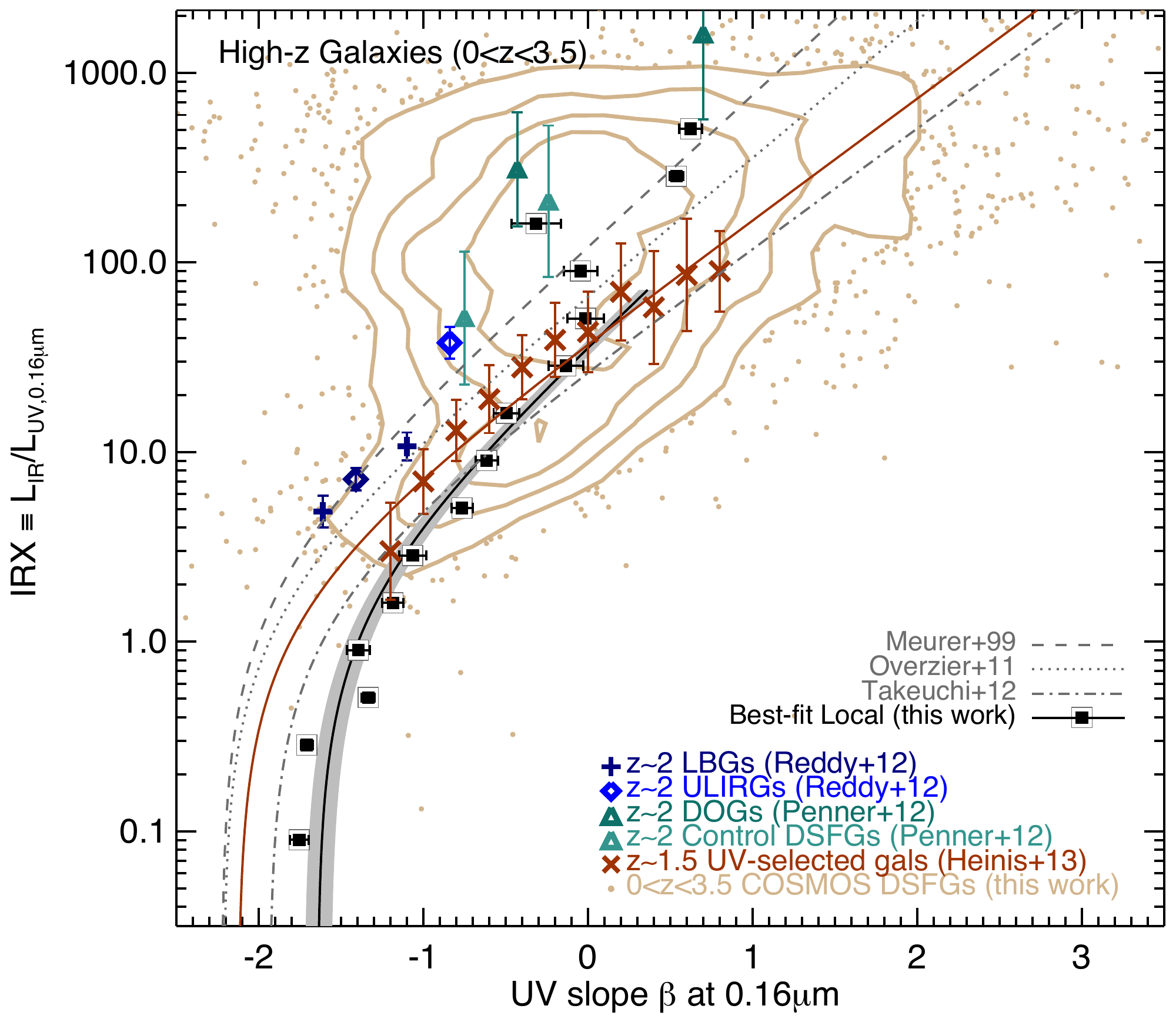}
\caption{\irx\ plotted against $\beta$, where $\irx\equiv$\iruv\ and
  $\beta$ is the rest-frame UV slope, for local galaxies (left panel)
  and for $z>0$ galaxies (right panel).  In both panels we overlay the
  original relation from \citet{meurer99a} for blue compact starbursts
  observed with {\it IUE} and {\it IRAS} (dashed gray line), the local
  Lyman-break analog (LBA) relation determined in \citet{overzier11a},
  which comprises a larger sample (dotted gray line). The revised
  relation to the \citet{meurer99a} galaxies from \citet{takeuchi12a}
  includes the aperture effects between {\it IUE} and \galex, where
  the latter correctly accounts for the integrated UV luminosity.  On
  the left panel we plot the local sample of 1034 nearby galaxies
  described by \citet{gil-de-paz07a} as well as 135 local (U)LIRGs
  from \citet{howell10a} with $\beta$ and \irx\ re-computed in this
  paper.  This local sample comprises a representative sample of all
  galaxies in the local volume; all galaxies are color-coded by their
  star formation rates from $\sim$0.03--100\sfr.  Our best-fit
  attenuation curve, irrespective of galaxy type, is shown in solid
  black with gray uncertainty envelope (given by
  Eq.~\ref{eq:irxbeta}).  Black square points denote the median UV
  slope in fixed \irx\ bins.  On the right panel we copy our
  measurement of the local relation and overplot several measurements
  of \irxb\ at high-redshift in galaxy samples that have direct-FIR
  measurements for $L_{\rm IR}$.  A stack of LBGs from
  \citet{reddy12a} is shown with a navy diamond, split into two
  $\beta$ bins (navy crosses), and low-luminosity $z\sim2$ ULIRGs
  (blue diamond).  Dust-Obscured Galaxies (DOGs), summarized in
  \citet{penner12a}, are shown in dark teal, while a comparable
  ULIRG-type control population is shown in lighter teal.  Burnt
  orange X's denote LBG stacking results in HerMES fields from
  \citet{heinis13a}.  The direct-detection COSMOS DSFGs in this work,
  spanning redshifts $0<z<5$, mostly at $z<2$, are shown in tan
  contours with individual outliers overplotted.}
\label{fig:irxbeta}
\end{figure*}

We compute the rest-frame UV luminosity of DSFGs by interpolating
their measured photometry to rest-frame 1600\,\AA\ as an AB apparent
magnitude, $m_{\rm 1600}$, with
\begin{equation}
L_{\rm UV} = \frac{4\pi\,D_{\rm L}^2\nu_{\rm 1600}}{(1+z)} \times 10^{-(48.60 + m_{\rm 1600})/2.5},
\label{eq:uv}
\end{equation}
where $D_{L}$ is the luminosity distance at redshift $z$ and $L_{\rm
  1600}$ is given in erg\,s$^{-1}$.
At $z\approx0$, UV luminosity is the interpolation of FUV and NUV
luminosity to 1600\,\AA\ (which is approximately equal to $L_{\rm
  FUV}$ given the FUV filter's proximity to rest-frame
1600\,\AA). Note that as the redshifts of the galaxies in the sample
increase, different observed bands are representative of the
rest-frame UV: the NUV {\it GALEX} 0.231\um\ filter at $z\simlt0.6$,
the Subaru $U$-band 0.346\um\ filter at $0.6\simlt z\simlt1.5$, and
the Subaru $B$-band 0.460\,\um\ filter at $1.5\simlt z\simlt2$, and so
on.  

We then compute the rest-frame UV slope, $\beta$, using a powerlaw
fit, where $\beta$ is defined by the relationship between flux and
wavelength such that $F(\lambda)\propto\lambda^{\beta}$ (where flux is
given in units of erg\,s$^{-1}$\,cm$^{-2}$\,\AA$^{-1}$).  We calculate
$\beta$ and the uncertainty on $\beta$ based on the multiple
photometric measurements in the rest-frame UV regime available in
COSMOS, namely 1230--3200\,\AA\ \citep[][ this range is chosen to
  primarily consist of UV radiation from young O and B stars while
  also being wide enough to make the measurement of $\beta$ feasible
  using multiple photometric
  bands]{calzetti94a,meurer99a,calzetti01a}.  Depending on redshift,
$\beta$ is calculated with 2--5 photometric bands, with an average of
3.3 bands per source.  To determine if any systematic bias is
introduced in the estimation of $\beta$ as a function of redshift
(thus, as a function of the different broad and narrow bands
available), we artificially redshift several different galaxy
templates \citep*[from][]{bruzual03a}, convolve the templates with the
available filter profiles, measure $\beta$ and compare to the
intrinsic rest-frame UV slope. While we infer an intrinsic scatter in
$\beta$ of up to 0.3 based on filter combination and noise, we infer no
systematic variation.

Note that since the measurement of $\beta$ is done photometrically, it
has the potential to be contaminated by stellar or interstellar
absorption features and also enhanced extinction around the rest-frame
2175\,\AA\ dust feature seen in high-metallicity environments like the
Milky Way \citep[but is less pronounced in low-metallicity
  environments like the LMC and SMC;][]{calzetti94a,gordon03a}.  This
2175\,\AA\ absorption feature is now known to exist in both low-$z$
galaxies \citep{conroy10a,wild11a} and some high-$z$ galaxies
\citep{buat11a,buat12a,kriek13a}; unfortunately, the intrinsic
attenuation curve in highly dusty galaxies is more difficult to
constrain since fewer rest-frame UV photons escape.  We do not think
this substantially affects our results if the galaxies we study follow
a Calzetti attenuation law, but we emphasize that further
work$-$especially on the most extreme starbursts$-$is necessary to
establish an empirical constraint on the relationship of attenuation
to metallicity.

Another potential bias introduced by performing this calculation
photometrically, is possible contamination from emission lines,
primarily \lya.  If a photometric redshift is slightly lower than a
galaxy's intrinsic redshift, emission from \lya\ might boost the
perceived rest-frame FUV emission indicating a bluer UV slope.  To
test for this contamination, we compare the best-fit $\beta$ from the
1230--3200\,\AA\ range with a more restrictive 2000--3200\,\AA\ range,
which should be substantially less affected by \lya\ contamination for
galaxies where photometric redshifts are better than $\Delta z=0.5$.
We find no systematic difference between the two different fits to
$\beta$ and therefore argue that emission line contamination of
$\beta$ in our sample is negligible and/or unlikely.

We have corrected for Galactic extinction effects on both UV
luminosity and $\beta$ using the dust maps of \citet*{schlegel98a},
with updates provided by \citet{peek10a}; this is of particular
importance for the nearby sample, which is distributed across the sky.
We use the Milky Way attenuation curve
\citep[$A_{\lambda}/A_V$ from][]{gordon03a} and the average extinction to
reddening relation at $V-$band of $A_V = 3.1 E(B-V)$ for the diffuse
Milky Way \citep{cardelli89a} to determine appropriate correction
factors $A_{\lambda}$ for observed bands from the FUV to z band.

\section{Analysis}\label{sec:analysis}

\subsection{\irxb\ for Nearby Galaxies}\label{sec:irxb_local}

The left panel of Figure~\ref{fig:irxbeta} shows the \irxb\ relation
in the local \citet{gil-de-paz07a} and \citet{howell10a} samples after
re-measuring $L_{\rm IR}$, $L_{\rm UV}$, and $\beta$ uniformly as
described in \S~\ref{sec:calculations}.  Galaxies' star formation
rates are denoted with color, ranging from blue to red, roughly
increasing monotonically with \irx.  It is clear that of the three
calibrations of this relationship in the local Universe, that of
\citet{takeuchi12a} provides the best fit, with both the widely-used
\citet{meurer99a} and \citet{overzier11a} fits offset in $\beta$ by
$\sim$0.7 towards the blue.  Again, this disagreement arises from the
corrected, integrated aperture of \galex\ between the
\citet{meurer99a} and \citet{takeuchi12a} works.  While
\citet{overzier11a} {\it do} correct for aperture effects, they
include a sample of particularly blue Lyman-break Analogs (LBAs) at
low redshift which causes the median fit to be substantially bluer
than that of \citet{takeuchi12a}.  All three prior literature works
are blueward of our fit at low \irx\ because of the subset of galaxies
used to derive the fit: starbursts.  Our inclusion of the
\citet{howell10a} LIRG data, and the full \citet{gil-de-paz07a}
heterogeneous sample spans a larger dynamic range in star formation
rates.  If we exclude LIRGs and low SFR systems, we recover a
shallower, bluer slope.

To determine the best-fit attenuation curve for local galaxies, we
first bin up our data in \irx\ intervals of 0.25\,dex.  We choose to
bin in \irx\ instead of $\beta$, as has been done elsewhere in the
literature, to avoid degeneracies at $\beta<-1.5$ (where a wide range
of \irx\ values all correspond to the same $\beta$) and at $\beta>0.5$
(where dusty galaxies become non-negligible with high values of
\irx\ for the same $\beta$).  We note that binning in $\beta$ or
\irx\ produces consistent results between $-1.4<\beta<0$ and
$1<\irx<50$.  Since these data are notably non-Gaussian in $\beta$ at
a given \irx, we compute a representative `mode' $\beta$ value by
measuring the peak of the $\beta$ histogram for the given \irx\ bin;
the black squares on Figure~\ref{fig:irxbeta} are these mode values.
Errors on the mode are determined by bootstrapping.  We exclude
galaxies with \irx$>60$ from the fit, where there is clear deviation,
a topic we will return to in subsequent sections.  A fit including the
data at \irx$>60$ is very shallow and a poor solution to all data
above \irx$>1$; we determined the cutoff of \irx$>60$ iteratively by
excluding the highest \irx\ bins until the reduced-$\chi^2$ was below
1.5.  Note that the galaxies at very high-$\beta$ and low \irx\ are
too few to affect the average binning or fit.
 We determine that
\begin{equation}
IRX = 1.68\times[10^{0.4[(3.36\pm0.10)+(2.04\pm0.08)\beta]}-1]
\label{eq:irxbeta}
\end{equation}
provides a good fit to the local data, with characteristic spread in
$\beta$ of 0.59.  This spread represents the standard deviation in the
difference of measured $\beta$ values from the expected $\beta$ value
(at a given \irx\ and given Eq.~\ref{eq:irxbeta}).  Here the
rest-frame UV extinction in magnitudes is represented by the quantity
A$_{UV}=3.36+2.04\beta$.  The factor of 1.68 accounts for the
bolometric correction of the original 40--120\,\um\ FIR studies to
total IR luminosities ($L_{\rm IR}$ integrated from 8 to 1000\,\um).
One can verify that 1.68 is an appropriate choice of bolometric
correction by generating SEDs with a variety of temperatures
\citep[e.g. by using the fitting method of][]{casey12a} and comparing
the integral luminosities between the two wavelength ranges.
Regardless, the exact value of the bolometric correction only is
relevant for readers wishing to compare our best-fit $A_{UV}$ to other
works that use similar notation.

Very dusty systems with $\irx>60$ are unexpectedly blue. Beyond
$\irx>60$, the mode value of UV color $\beta$ hits a wall at
$\beta\approx-0.1$.  We note that this result is slightly different
from the analysis of the local sample presented in
\citet{overzier11a}, who only measured galaxies with $\irx\simgt100$
as slight outliers; the difference is primarily due to the redder
colors we measure in the low luminosity, low-$\irx$ systems than any
discrepancy at the high-$\irx$ end.  Also note that, while bluer than
expected from the \irxb\ relation, these very dusty galaxies are not
as blue as some Lyman-Break Galaxies in the early Universe.  For the
rest of our analysis, unless stated otherwise, whenever we refer to
the `local' or `nominal' \irxb\ relation, we are explicitly referring
to the relationship derived here, represented by
Equation~\ref{eq:irxbeta}.

\subsection{Context of \irxb\ at high-$z$}

\begin{figure*}
\centering
\includegraphics[width=1.99\columnwidth]{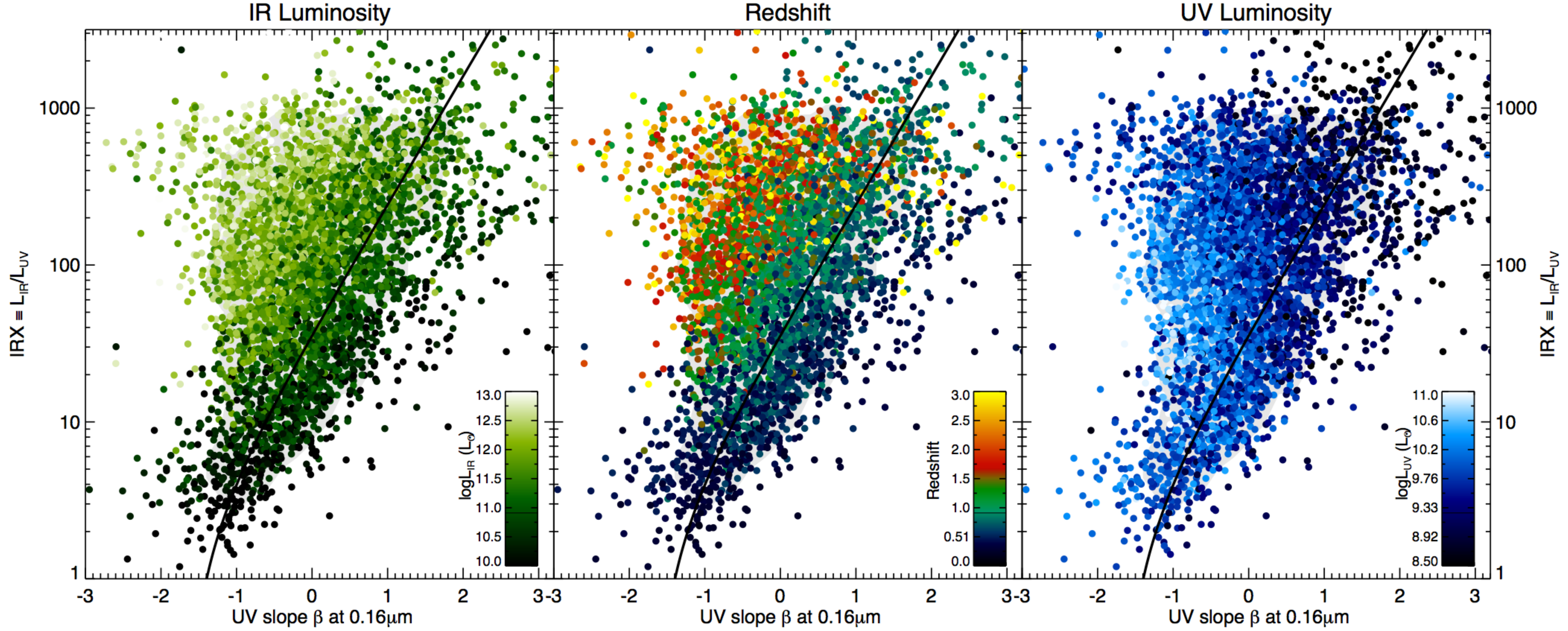}
\caption{Three panels illustrating the \irxb\ relation with respect to
  DSFGs' IR luminosity (left), redshift (middle), and UV luminosity
  (right), as indicated by point color.  The nominal \irxb\ relation as
  given in Equation~\ref{eq:irxbeta} is shown in solid black on each
  panel for context. 
  The higher a galaxy's IR luminosity or redshift, the farther away
  that galaxy sits from the \irxb\ relation towards bluer rest-frame
  UV colors, i.e. the steepest gradients in IR luminosity and redshift
  are orthogonal to the \irxb\ relation.  On the other hand, the
  steepest gradient in UV luminosity is parallel to the \irxb\ relation.
 }
\label{fig:irxbeta_grad}
\end{figure*}

The right panel of Figure~\ref{fig:irxbeta} places the local
\irxb\ relationship in context at higher redshift.  The local relations,
including the best-fit relation in Equation~\ref{eq:irxbeta} are
overplotted with some comparative data-sets from the literature, all
of which involve direct far-infrared and direct rest-frame UV
measurements (other works using indirect methods have been omitted).
We include the 114 LBG stacking result from \citet{reddy12a}, whose
measurement sits auspiciously along the \citet{meurer99a} local
attenuation curve, and their 12-galaxy, low-luminosity ULIRG
comparison sample, which sits at slightly elevated \irx\ relative to
\citeauthor{meurer99a}  We also include the IR-stacking results on
38000$+$ UV-selected $z\sim1.5$ galaxies from \citet{heinis13a}.  Note
that the \citeauthor{heinis13a} data agree remarkably well with our
locally derived \irxb\ relation and not the \citeauthor{reddy12a}
stacking results.  We attribute this disagreement to the fact that the
\citeauthor{heinis13a} sample is not color selected.  One puzzling
aspect of the \citeauthor{heinis13a} sample is the continuation of the
relation to very red colors, $\beta\approx1$.  What is different in
these galaxies at $z\sim1.5$ that allows for this `extra' reddening?
We believe sample selection and binning on $\beta$ rather than $\irx$
(in this case out of necessity since the galaxies are not directly
detected in the infrared). The \citeauthor{heinis13a} sample does not
include galaxies explicitly {\it selected} by their dust emission.

As we found with our local sample, an exclusion of galaxies selected
at IR wavelengths skews the \irxb\ relation towards redder colors
\citep[consistent with what was found by][]{takeuchi12a}.  As such, we
do not attempt to derive a best-fit $z>0$ \irxb\ relation,
since it is quite clear from Figure~\ref{fig:irxbeta} (right panel)
that most DSFGs do not follow a strict \irxb\ relationship.  This
would only be possible if we had a larger dynamic range in \irx,
similar to what is available for the local sample.

It is quite clear that sample selection impacts the interpretation of
\irxb.  In relation to the UV-selected samples, the aggregate
\irx\ and $\beta$ values presented for dusty galaxies in
\citet{penner12a} are notably offset in a similar manner as the local
LIRGs and ULIRGs of \citet{howell10a}, emphasizing that dusty galaxies
are bluer than the nominal \irxb\ relation would predict.  Where do
the COSMOS DSFGs lie in relation to these other $z>0$ measurements?
The tan background contours on the right panel of
Figure~\ref{fig:irxbeta} represent the COSMOS DSFGs. They range from
being directly on, or even below, the local relation to being above
the relation by $\sim2\,$dex.  In the next subsection we investigate
the difference between DSFGs which lie on the relation and those
sitting substantially above.

DSFG characteristics in the $IRX-\beta$ plane show strong migration
with infrared luminosity, ultraviolet luminosity and redshift.  The
infrared luminosity and redshift migration goes hand-in-hand, whereby
sources at the highest redshifts have the highest IR luminosities,
based on our DSFG IR flux density based selection.  Indeed,
Figure~\ref{fig:irxbeta_grad} shows that contours of equal $L_{\rm
  IR}$ and $z$ roughly trace one another.  On the other hand, contours
of constant UV luminosity are orthogonal to those of constant $L_{\rm
  IR}$. This is expected given our IR sample selection and,
importantly, assuming that there is only loose correlation between
$L_{\rm UV}$ and $L_{\rm IR}$.  From this plot, we can clearly see
that galaxies with the highest IR luminosities and redshifts lie
farthest away from the local attenuation curves in the \irxb\ plane.
What does this tell us physically?
To understand the underlying physical conditions leading to this
$L_{\rm IR}$ or $z$-driven migration of DSFGs in the \irxb\ plane, we
first have to ensure that this is not caused by sample selection effects.

\subsection{Testing for Selection Bias}

The COSMOS field where the DSFGs are selected has the benefit of very
deep ancillary data from the UV through the near-infrared for use in
fitting stellar emission \citep{capak07a}.  To test for selection
biases in the \irxb\ plane, we first complete an analysis of the depth
of coverage in every rest-frame UV filter and rest-frame IR filter
used to calculate $L_{\rm UV}$, $\beta$ and $L_{\rm IR}$.  In the
rest-frame UV, we determined completeness as a function of magnitude
for all galaxies in the field for each filter, using the photometry 
reported by \citet{ilbert10a}.  In the far-infrared, we know the
characteristic depth of {\it Herschel} coverage at 100\,\um, 160\,\um,
250\,\um, 350\,\um\ and 500\,\um\ as reported by \citet{lutz11a} and
\citet{nguyen10a} and how to apply deboosting factors as appropriate
to estimate intrinsic flux densities from raw, corrected for confusion
noise \citep{roseboom10a,roseboom12a}. This single characteristic flux
limit is primarily blurred by the uncertainty in deboosting factors
(and to a much lesser extent, variations in instrument noise).

\begin{figure}
\begin{center}
\includegraphics[width=0.9\columnwidth]{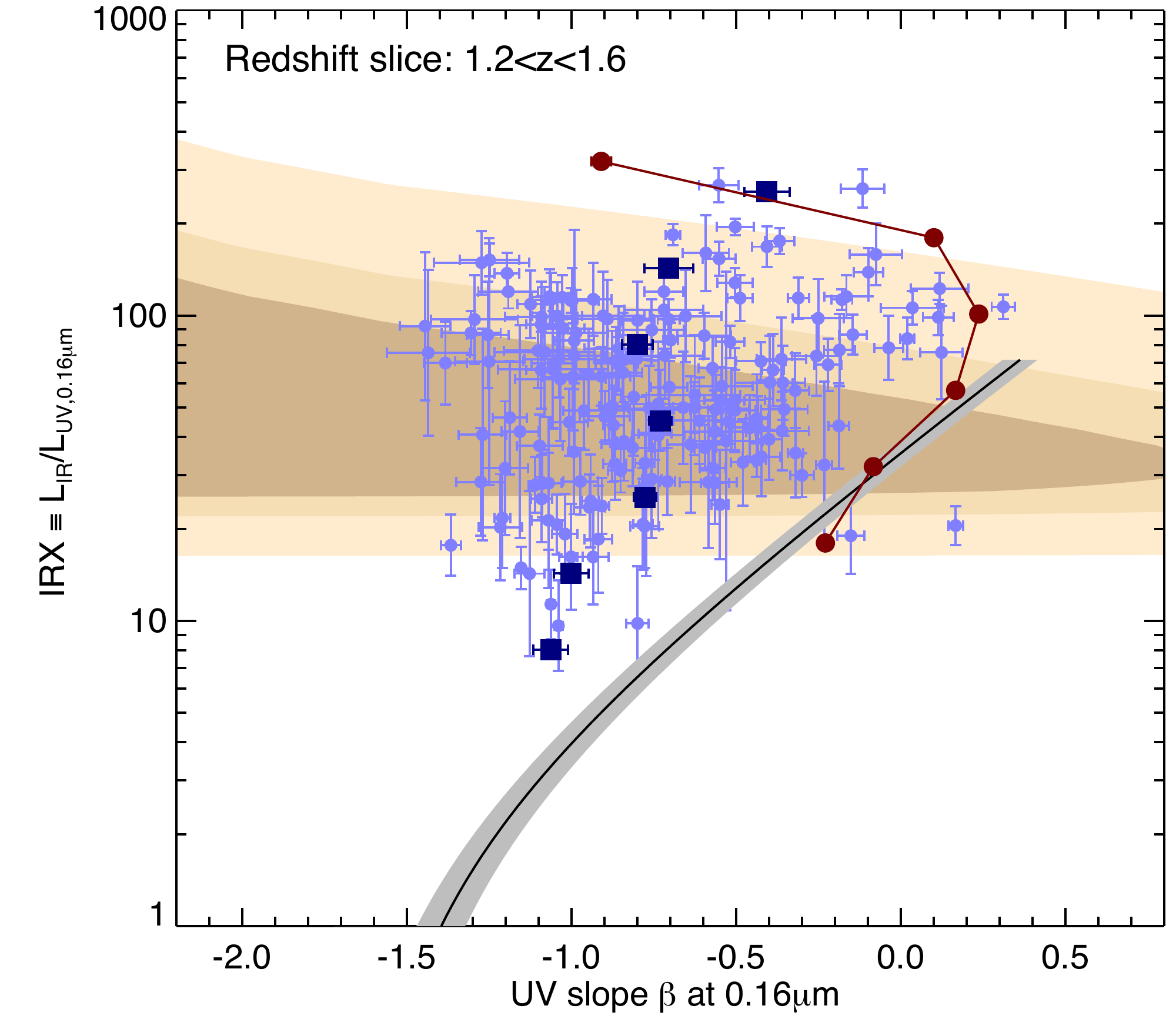}
\includegraphics[width=0.9\columnwidth]{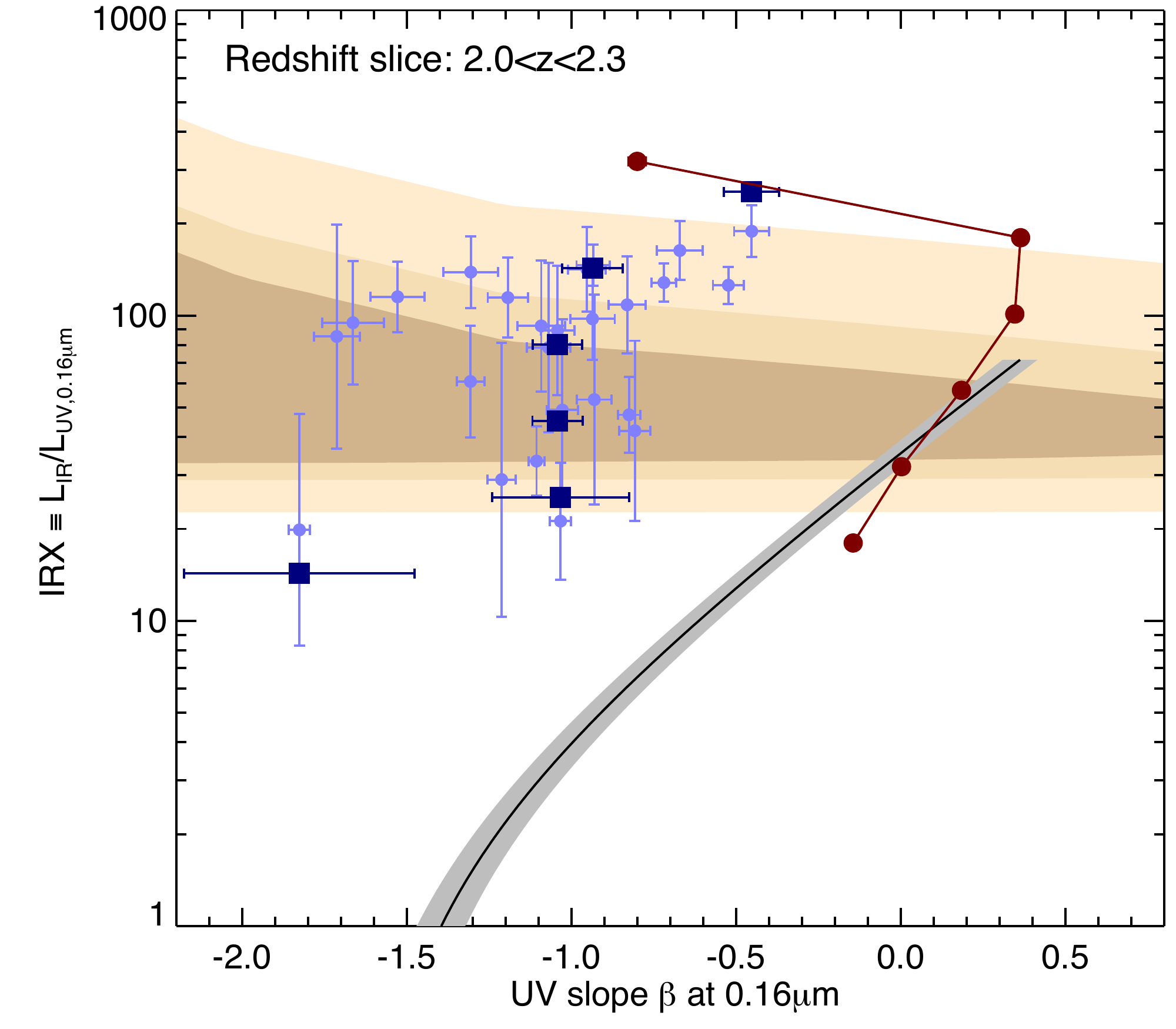}
\end{center}
\caption{An illustration of the selection effects within our DSFG
  sample in two redshift slices.  Light blue points are individual
  DSFGs that fall in the given redshift bins, and dark blue square points are
  the median values of $\beta$ for DSFGs falling in the given
  \irx\ bins.  Tan shaded areas isolate accessible \irxb\ space where
  sources are over 95\%, 90\%\ and 80\%\ likely to be included in our
  survey.  The limit at low \irx\ is caused by the lower limit for
  direct IR detection combined with an upper limit on $L_{\rm UV}$.
  The limit at high-\irx, which is a shallow function of $\beta$,
  depends on the depth of the UV/optical bands in COSMOS.  We overplot
  the local relation from Figure~\ref{fig:irxbeta} in black (gray
  uncertainty) and simulate the selection functions impact on the
  local relation (round red points).  This demonstrates that the offset
  towards lower $\beta$ in DSFGs is not driven by selection effects.
}
\label{fig:selection}
\end{figure}

To tease out the possible selection biases in the \irxb\ plane, we
construct a grid across $-3<\beta<3$ and $-2<\log(IRX)<4$, with a
binsize of 0.05 in both quantities.  Similarly, we construct a
parallel grid of plausible IR luminosities and redshifts, which
reaches far beyond the range of our observed sample: $6<\log(L_{\rm
  IR})<14$, with a binsize of 0.05 and $0<z<5$ with $\Delta
z$=0.01. For a given $\beta$ and \irx, at a given $L_{\rm IR}$ and
$z$ we compute $L_{\rm UV}$ (where $L_{\rm UV}\,=\,L_{\rm IR}/IRX$),
rest-frame magnitude $m_{1600}$ at 1600\,\AA\ via Eq~\ref{eq:uv}, and
assuming $F\propto\lambda^\beta$, we compute the observed
AB-magnitudes in the COSMOS filters that span the rest-frame UV at
the given redshift $z$.  We then add statistical noise (instrumental
and confusion) to the measured magnitudes to be consistent with real
data noise, remeasure $\beta$, and determine the probability that the
given source would be detectable in our survey given the completeness
curves described in the prior paragraph.  This test is then repeated
across the entire grid of possible $\beta$, \irx, $L_{\rm IR}$ and $z$
values\footnote{As an aside, our tests indicate there is no systematic
  bias in measuring $\beta$ at low luminosities.}, measuring a
probability of inclusion in our DSFG sample for each permutation.

To complete this test of selection bias, we must also assign a
probability of detection based on redshift and infrared luminosity
(two of the four independent parameters in the above test for
rest-frame UV detectability).  This requires translating our
\herschel\ flux density limits into luminosity, which requires some
assumption on IR SED characteristics, primarily dust temperature.
\citet{symeonidis13a} and \citet{lee13a} both present an observed
trend of increasing dust temperature with luminosity which is {\it
  unbiased} with respect to temperature, unlike similarly luminous
systems selected at 850\,\um\ \citep{blain04a,chapman04a,casey09b}.
This relation is also observed locally, but shifted towards slightly
warmer temperatures, perhaps indicative of size differences in the
population, local galaxies being more compact \citep{swinbank13a}.
This dust temperature shift at high redshifts is explicitly
illustrated in figure~25 of \citet*{casey14a}.  We adopt a
representative SED shape model to characterize the dust temperature,
or SED peak wavelength, in terms of redshift and IR luminosity via
\begin{equation}
\langle\log\lambda_{\rm peak}\rangle = \eta [\log\!L_{\rm IR} - (\log\!L_{\rm 0} + \gamma\log(1+z))] + \mu.
\end{equation}
At $z=0$, this equation simplifies to $\langle\log\lambda_{\rm
  peak}\rangle=\eta(\log\!L_{\rm IR}-\log\!L_{0})+\mu$.  Here
$\eta=-0.062$ is the slope of the correlation, $\log\!L_{0}=10.60$
sets an arbitrary luminosity zero point at $z=0$, and $\mu=1.99$ is
the average $\log\lambda_{\rm peak}$ at that luminosity.  The redshift
evolution of $L_{0}$ is assumed to take the form $(1+z)^{\gamma}$, as
is often used to model the evolution of $L_\ast$ in luminosity
functions \citep[e.g.][]{caputi07a,goto10a}, with best-fit
$\gamma=2.7$.  This function has characteristic scatter around
$\log\lambda_{\rm peak}$ of 0.045 (Casey \etal\ in prep.).  This model
gives us a reliable and realistic estimate for the limiting luminosity
of our \herschel\ data.  Note, however, that the temperature-redshift
dependence of this model only weakly impacts our conclusions regarding
detectability in this paper, and only at the low \irx\ end, and is not
a function of the rest-frame UV slope $\beta$.

Figure~\ref{fig:selection} offers an illustration of our survey's
selection effects in the \irxb\ plane in two redshift slices.  Areas
shaded in tan are likely to be covered by the detection limits of
COSMOS data.  To answer the question of whether or not our deviation
towards bluer colors is driven by selection effects, we model a
population of 10$^6$ galaxies which cluster about the local
\irxb\ curve (from Eq.~\ref{eq:irxbeta}) with measured scatter.  After
applying our selection limits to that sample and binning in \irx, we
are left with the rounded red points in Figure~\ref{fig:selection},
which are significantly more red than the observed samples (shown in
blue).  This leads us to conclude that selection effects do not drive
this observed blue $\beta$ offset in DSFGs.

\subsection{Comparison to Literature IR-selected Galaxies}

Our result that high-SFR, high-$L_{\rm IR}$ galaxies are much bluer
than they would nominally be expected to be\footnote{Where the
  `expected' $\beta$ for a given \irx\ would be described by our
  Equation~\ref{eq:irxbeta}.} is consistent with previous analyses on
smaller case studies of luminous infrared galaxies, both at low-$z$
\citep{goldader02a,howell10a} and high-$z$.
The offset from the \irxb\ relation was attributed in
\citet{goldader02a} to the significant spatial disassociation of the
UV-luminous and IR-luminous portions of each galaxy.
\citeauthor{goldader02a} speculated (at a time when very little was
known about high-$z$ DSFGs) that local ULIRGs might be unusual in that
their physical compactness makes it difficult for much rest-frame UV
light to escape (i.e. at a given UV slope $\beta$, a local ULIRG will
have a higher \iruv\ ratio than a normal star-forming galaxy due to
its compactness) and at high-$z$, this might not be problematic if
DSFGs are more spatially extended, as has often been found to be the case
\citep[e.g.][]{ivison98a,ivison10c,ivison11a,hodge12a}.

The submillimeter galaxy population \citep[SMGs;][]{smail97a} were,
like the local ULIRGs, found to be `bluer' than expected given the
incredible \iruv\ ratios present, \simgt100 \citep{smail04a}.
\citet{chapman05a} pointed out that this implied that UV-based star
formation rates (SFRs) of SMGs underestimated their total SFRs by
factors as large as $\sim$120 even {\it after} the UV SFRs were corrected
for extinction \citep[cf.][]{adelberger00a}.
\citet{penner12a} also found that dusty galaxies$-$even when
they are pre-selected to be incredibly red at optical
wavelengths$-$have rest-frame UV characteristics bluer than expected
given the local attenuation curve.  Furthermore, a comparison control
sample of IR-selected galaxies that are not DOGs revealed a similar,
yet less extreme result, completely in line with expectation given our
results.

\begin{figure}
\includegraphics[width=0.99\columnwidth]{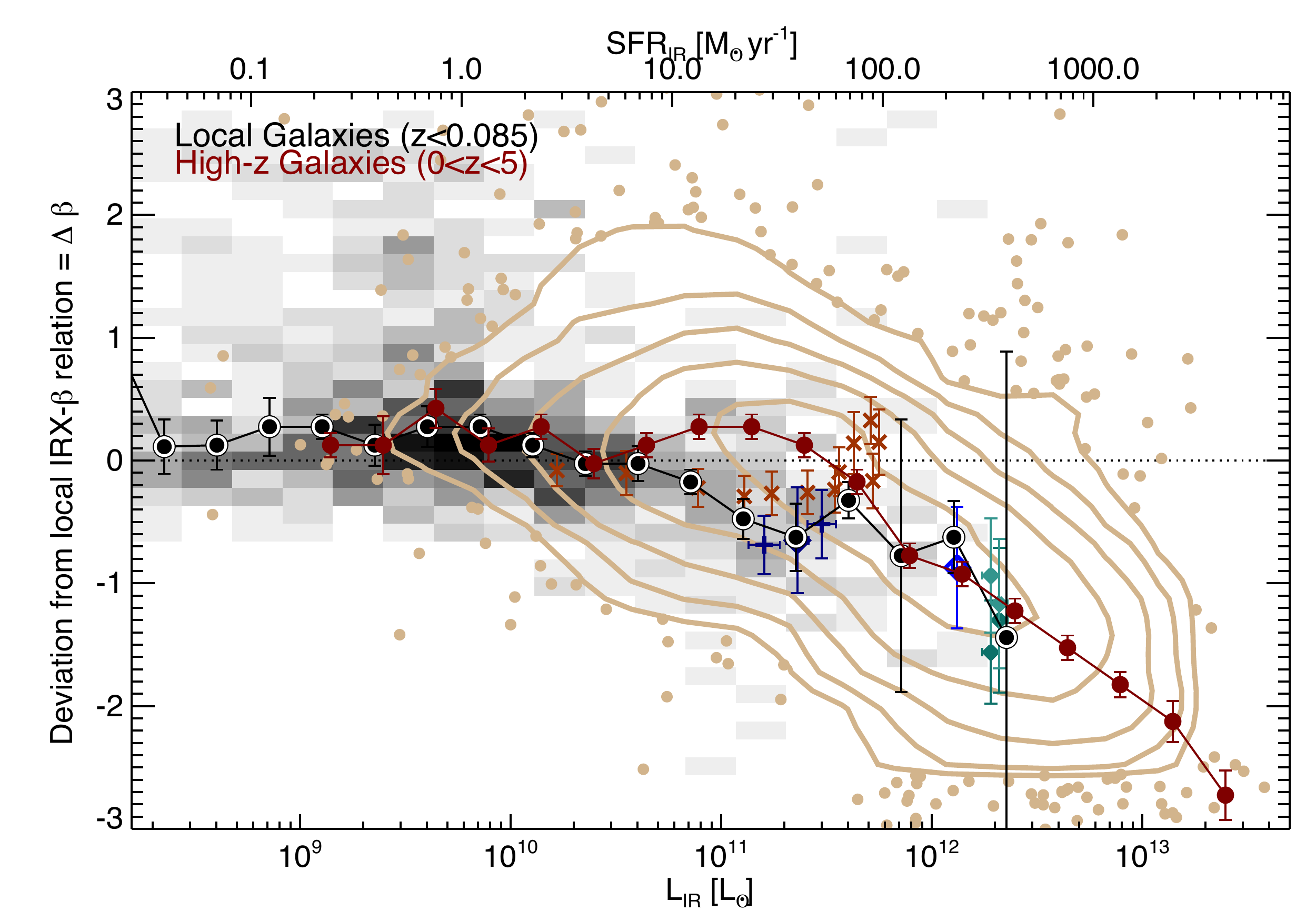}
\includegraphics[width=0.99\columnwidth]{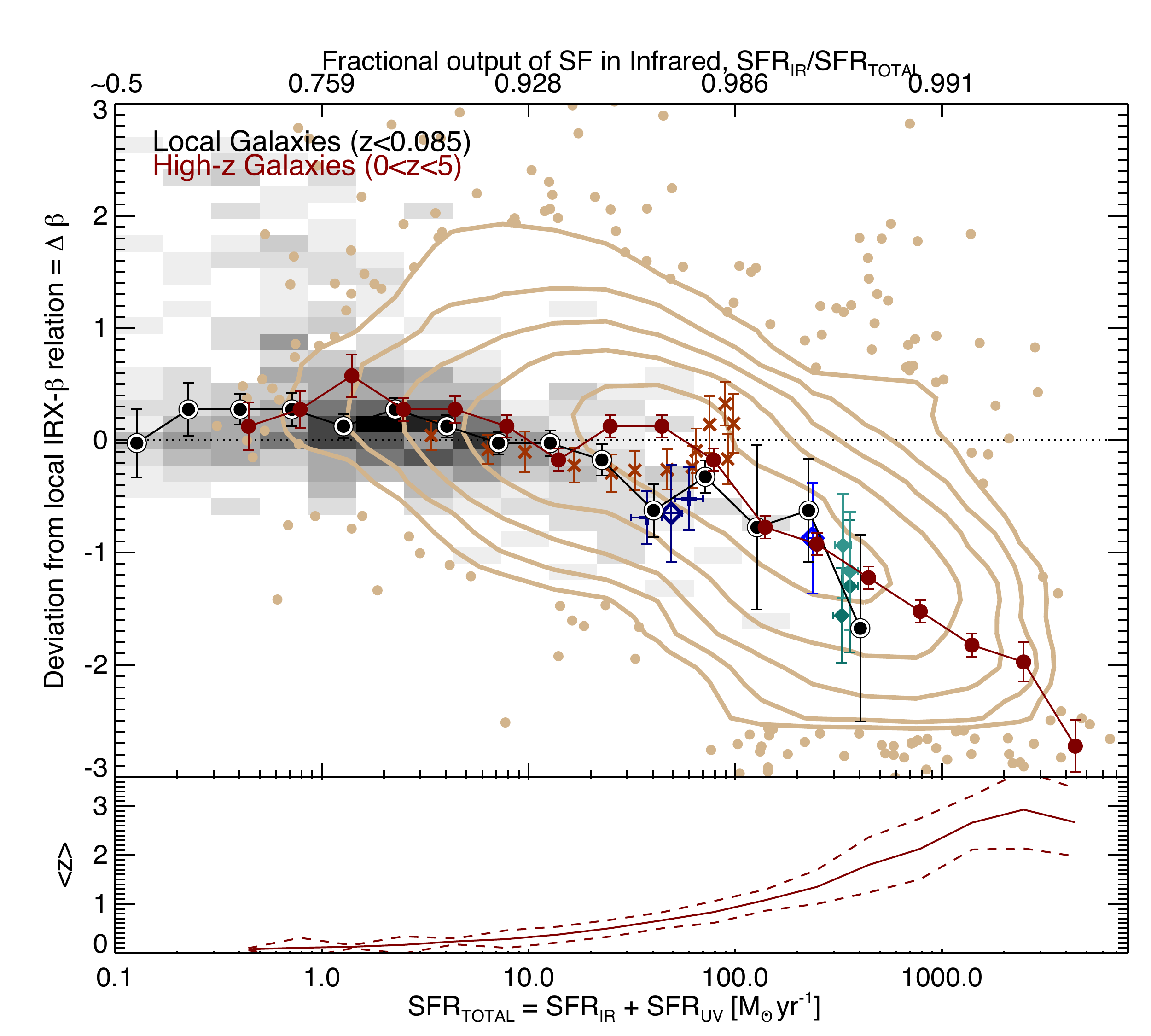}
\caption{The deviation from the nominal \irxb\ relation, measured as a
  difference in rest-frame UV slope, $\Delta\beta\equiv \beta_{\rm i}
  - \beta_{\rm exp}$, where $\beta_{\rm exp}$ relates to \irx\ via
  Equation~\ref{eq:irxbeta}, against IR luminosity (top) and total
  star formation rate (bottom).  Here we use SFR$_{\rm IR}$+SFR$_{\rm
    UV}$ as a proxy for total star formation rate, which above
  $\simgt$10\,\sfr\ has 90\%\ of its energy output in the IR (see top
  axis of bottom plot).  The local samples are shown as a shaded gray
  backdrop and overplotted black median values, while the $z>0$
  COSMOS DSFGs are contoured in tan with median values in dark red.
  The median redshift of the sample at a given SFR is shown in the
  bottom inset ranging from $0<z<3$.  Literature studies with
  published $L_{\rm IR}$ values are overplotted with the same symbols
  as in Figure~\ref{fig:irxbeta} (right). In both diagrams, we see a
  strong break at $L_{\rm IR}\approx10^{11.5}$\lsun\ or
  SFR$\approx40\,$\msun\,yr$^{-1}$, above which galaxies of all epochs
  are bluer than expectation.}
\label{fig:dbeta}
\end{figure}

\begin{figure}
\includegraphics[width=0.99\columnwidth]{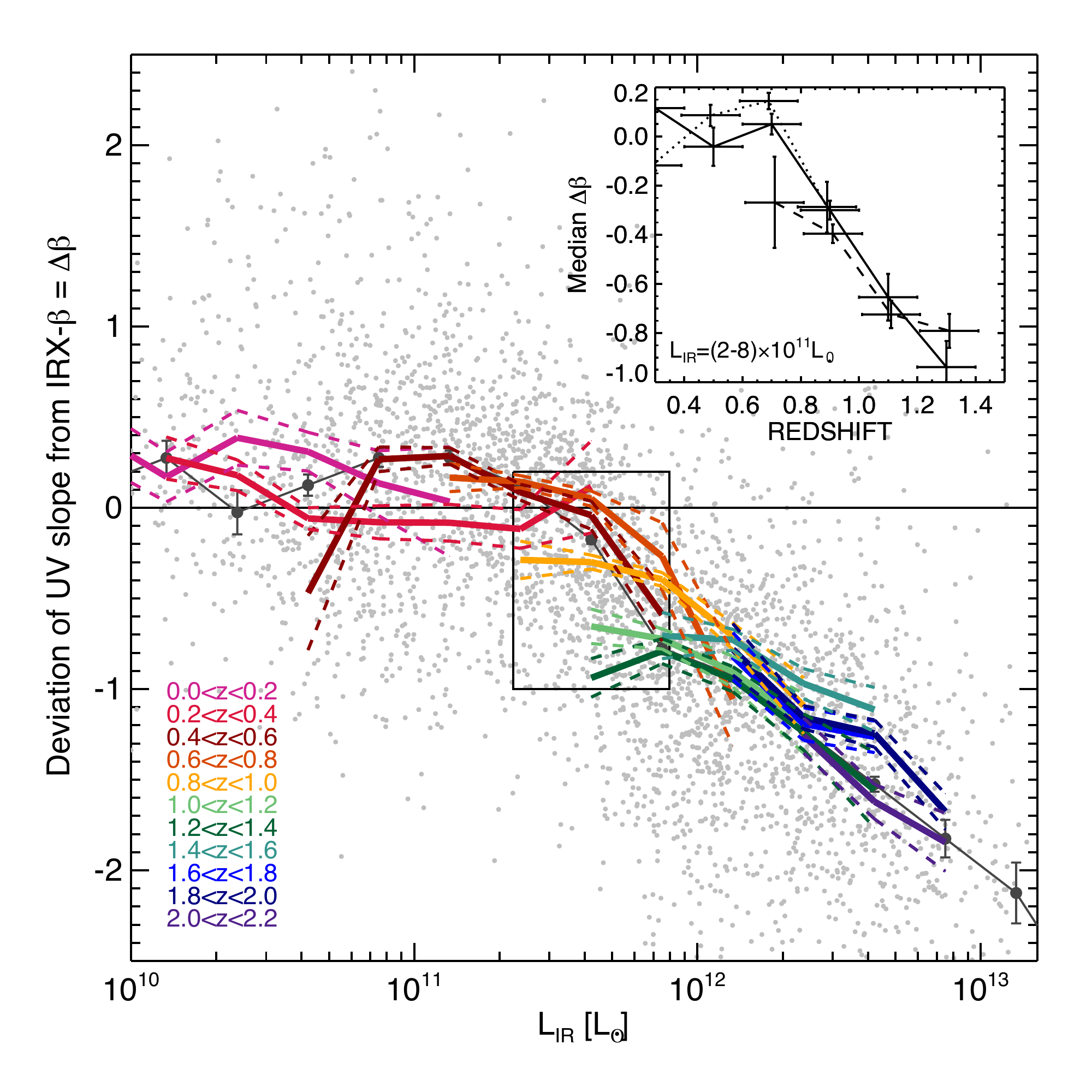}
\caption{Same as Figure~\ref{fig:dbeta} but split into $\Delta z=0.2$
  redshift bins to probe a possible underlying redshift evolution
  distinct from the overall luminosity-driven \irxb\ deviation.  The
  weighted mean $\Delta\beta$ value for a given redshift bin and
  luminosity bin is shown with a bootstrap-estimated uncertainty.  The
  only significant redshift evolution which is observed is seen
  between $0.6<z<1.4$ and $2-8\times10^{11}$\lsun\ (central box),
  whereby galaxies at lower redshift are redder by $\Delta\beta=1$.
  Galaxies at $z<0.6$ and $L_{\rm IR}<2\times10^{11}$\lsun\ are too
  few in number to measure evolution, while evolution is simply not
  seen in the higher luminosity bins above $z>1.4$.
 The inset plot shows the change in $\Delta\beta$ with redshift within
 this interval for the three luminosity bins at
 2.4$\times$10$^{11}$\lsun\ (dotted),
 4.2$\times$10$^{11}$\lsun\ (solid), and
 7.5$\times$10$^{11}$\lsun\ (dashed).  }
\label{fig:dbetaz}
\end{figure}
\subsection{Deviation from the \irxb\ relation}

From Figure~\ref{fig:irxbeta_grad} we gather that the deviation from
the nominal \irxb\ relation is either a result of IR luminosity or
redshift. Because local ULIRGs also deviate from \irxb, we attribute
the correlation of galaxies' deviation to their IR luminosities.
Figure~\ref{fig:dbeta} investigates this deviation as a function of
$L_{\rm IR}$.
Due to the degeneracy of \irx\ at low $\beta$, we measure the
deviation from the nominal \irxb\ relation (that given by
Eq.~\ref{eq:irxbeta}) as the difference in UV-slope, or $\Delta\beta$.
\citet{howell10a} present a similar plot in terms of $\Delta$\irx for
the local sample (black/gray shaded background on our plot).  Although
there is large scatter in $\Delta\beta$, owing to the complex star
formation histories in a heterogeneous galaxy population, both the
local and $z>0$ samples share a characteristic `break'
luminosity, above which galaxies systematically deviate from
\irxb\ towards bluer UV slopes.  The strength of the deviation
increases with increasing luminosity.  The `break' luminosity for the
local sample appears to sit at $\approx$10$^{11-11.5}$\lsun\ while the
break in the COSMOS DSFG sample lies clearly at
$\approx10^{11.5}$\lsun.

Following \citet{kong04a} and some discussion presented in
\citet{reddy06a}, this deviation at high $L_{\rm IR}$ is anticipated.
Galaxies with more intense, more recent star formation will be
intrinsically bluer for a fixed dust attenuation, because the
underlying emission at UV wavelengths is dominated by a higher
proportion of young O stars contributing to the stellar continuum
emission at 1216--1600\,\AA.  In the dusty star-forming environments
of ULIRGs where attenuation is substantial, $L_{\rm IR}$ can be
directly mapped to the total SFR. The break in the $L_{\rm
  IR}-\Delta\beta$ plot correlates with increasing star formation
rate.  The bottom panel of Figure~\ref{fig:dbeta} investigates
$\Delta\beta$ as a direct function of SFR$_{\rm total}$, or SFR$_{\rm
  IR}+$SFR$_{\rm UV}$.  To emphasize the relative contributions of IR
and UV to SFR$_{\rm total}$, the top axis of the bottom panel
indicates the fractional output of star formation in the infrared.
Similar to the break IR luminosity quoted above, the break in SFR is
seen at $\approx30-50$\sfr\ in both local and $z>0$ samples
\citep[assuming a Salpeter IMF;][]{salpeter55a}.

Although the deviation from \irxb\ is clearly systematic above a given
$L_{\rm IR}$, the huge scatter of $\sigma_{\Delta\beta}=1$ leads us to
ask whether or not there is also any underlying redshift evolution.
 Figure~\ref{fig:dbetaz} breaks up the $L_{\rm
  IR}-\Delta\beta$ plot into redshift bins with $\Delta z=0.2$, where
the tracks in redshift are representative of the weighted mean and
uncertainties are bootstrapped.
Slight differences between median $\Delta\beta$ exist with redshift
and are seen most prominently in the half decade of $L_{\rm
  IR}\approx10^{11.5-12}$\lsun.  The two luminosity bins
at $4\times10^{11}$ and $7.5\times10^{11}$\lsun\ show substantial
evolution between $z=0.6$ and $z=1.4$.  Over the corresponding cosmic
time, the median UV color in galaxies of equal luminosity shifts by
$\Delta\beta=1$, i.e. it is substantially redder at lower redshifts.  At
higher luminosities (and also higher redshifts), no significant
differences are detected between epochs.

\begin{deluxetable*}{l@{ }c@{ }c@{ }c@{ }c@{ }c@{ }c@{ }c}
\tablecolumns{5}
\tablecaption{DSFG Contaminants to high-$z$ LBG searches}
\tablehead{
\colhead{Target}   & \colhead{Selection Criteria} & \multicolumn{5}{c}{\underline{\sc DSFG Contaminant Characteristics}} &  \colhead{\%} \\
\colhead{redshift} & \colhead{}                   & \colhead{$N_{\rm contam}$} & \colhead{$\rho_{\rm contam}$} & \colhead{Av. app.} & \colhead{$z_{\rm phot}$} & \colhead{Av. implied} & \colhead{LBG} \\
\colhead{}         & \colhead{}                   & \colhead{}                 & \colhead{[arcmin$^{-2}$]}     & \colhead{mag}  & \colhead{}               & \colhead{$\beta$} & \colhead{{\sc contam.}} \\}
\startdata
\multicolumn{8}{c}{{\sc Optical-Only LBG Dropout Selection}} \\
\hline
$z\sim2.5$ & {\footnotesize [$U_{\rm 300}-B_{\rm 450}>1.1$] $\wedge$ [$B_{\rm 450}-I_{\rm 814}<1.5$] $\wedge$}                & 15 & 2.4$\times$10$^{-3}$ & ($B$)26.7 & 2.7 & $-0.37\pm0.13$ & 0.01\%\ \\
           & {\footnotesize [$U_{\rm 300}-B_{\rm 450}>0.66$($B_{\rm 450}-I_{\rm 814}$)$+1.1$] $\wedge$ [$U>27$] }             &    &                      &        & \\
$z\sim4$   & {\footnotesize [$B_{\rm 435}-V_{\rm 606}>1.1$] $\wedge$ [$V_{\rm 606}-z_{\rm 850}<1.6$] $\wedge$}                & 35 & 5.5$\times$10$^{-3}$ & ($V$)26.3 & 2.9 & $-0.30\pm0.18$ & 0.37\%\ \\
           & {\footnotesize [$B_{\rm 435}-V_{\rm 606}>$($V_{\rm 606}-z_{\rm 850}$)$+1.1$] $\wedge$ [$UB>27$]}                 &    &                      &        & \\
$z\sim5$   & {\footnotesize [$V_{\rm 606}-i_{\rm 775}>1.2$] $\wedge$ [$i_{\rm 775}-z_{\rm 850}<0.6$] $\wedge$}                & 26 & 4.1$\times$10$^{-3}$ & ($i$)26.2 & 2.0 & $0.0\pm0.2$ & 3.2\%\ \\
           & {\footnotesize [($V_{\rm 606}-i_{\rm 775}>0.9$($i_{\rm 775}-z_{\rm 850}$))$\vee$($V_{\rm 606}-i_{\rm 775}>2$)] } &    &                      &        & \\
           & {\footnotesize $\wedge$ ($UBV>27$)}                                                                              &    &                      &        & \\
$z\sim6$   & {\footnotesize [$i_{\rm 775}-z_{\rm 850}>1.3$] $\wedge$ [[$z_{\rm 850}-J_{\rm 110}<0.6$] $\vee$}                 &  7 & 1.4$\times$10$^{-3}$ & ($z$)26.1 & 2.0 & $-0.4\pm0.2$ & 8.7\%\ \\
           & {\footnotesize [$i_{\rm 775}-z_{\rm 850}>2/3$($z_{\rm 850}-J_{\rm 110}$)$+1.02$]] $\wedge$ [$UBVi>27$]}          &    &                      & \\

\hline
\multicolumn{8}{c}{{\sc Optical and Near-IR LBG Dropout Selection}} \\
\hline
$z\sim4$   & {\footnotesize [$B_{\rm 435}-V_{\rm 606}>1$] $\wedge$ [$i_{\rm 775}-J_{\rm 125}<1$] $\wedge$}                    &  6 & 9.4$\times$10$^{-4}$ & ($V$)26.3 & 2.9 & $-0.30\pm0.18$  & $<$0.01\%\ \\
           & {\footnotesize [$B_{\rm 435}-V_{\rm 606}>1.6$($i_{\rm 775}-J_{\rm 125}$)$+1$] $\wedge$ [$UB>27$]}                &    &                      &              & \\
$z\sim5$   & {\footnotesize [$V_{\rm 606}-i_{\rm 775}>1.2$] $\wedge$ [$z_{\rm 850}-H_{\rm 160}<1.3$] $\wedge$}                &  2 & 3.2$\times$10$^{-4}$ & ($i$)26.2 & 2.0 & $0.0\pm0.2$ & 0.01\%\ \\
           & {\footnotesize [$V_{\rm 606}-i_{\rm 775}>0.8$($z_{\rm 850}-H_{\rm 160}$)$+1.2$] }                                &    &                      &         & \\
           & {\footnotesize $\wedge$ ($UBV>27$)}                                                                              &    &                      &         & \\
$z\sim6$   & {\footnotesize [$i_{\rm 775}-z_{\rm 850}>1.0$] $\wedge$ [$Y_{\rm 105}-H_{\rm 160}<0.45$] $\wedge$}               &  0 & $<$1.6$\times$10$^{-4}$ & ($z$)26.1 & 2.0 & $-0.4\pm0.2$ & $<$0.04\%\ \\
           & {\footnotesize [$i_{\rm 775}-z_{\rm 850}>0.777$($Y_{\rm 105}-H_{\rm 160}$)$+1.0$]] $\wedge$ [$UBVi>27$]}         &    &                      &        & \\
$z\sim7$   & {\footnotesize [$z_{\rm 850}-Y_{105}>0.7$] $\wedge$ [$J_{\rm 125}-H_{\rm 160}<0.45$] $\wedge$ }                  &  3 & 4.7$\times$10$^{-4}$ & ($Y$)24.9 & 1.9 & $-1.0\pm0.2$ & 0.40\%\ \\
           & {\footnotesize [$z_{\rm 850}-Y_{\rm 105}>0.8$($J_{\rm 125}-H_{\rm 160}$)$+0.7$] $\wedge$ [$UBViz>27$]}           &    &                      &        & \\
$z\sim8$   & {\footnotesize [$Y_{\rm 105}-J_{\rm 125}>0.45$] $\wedge$ [$J_{\rm 125}-H_{\rm 160}<0.5$] $\wedge$ }              &  1 & 1.6$\times$10$^{-4}$ & ($J$)24.1 & 4.3 & $-1.11\pm0.14$ & 0.34\%\ \\
           & {\footnotesize [$Y_{\rm 105}-J_{\rm 125}>0.75$($J_{\rm 125}-H_{\rm 160}$)$+0.525$]} $\wedge$ [$UBVizY>27$]       &    &                      &        & \\
$z\sim10$  & {\footnotesize [$J_{\rm 125}-H_{\rm 160}>1.2$] $\wedge$ [$H_{\rm 160}-$[$3.6$]$<1.4$ $\vee$}                     &  0 & 3.2$\times$10$^{-4}$ & ($H$)23.0 & 2.8 & $-0.2\pm0.4$ & $<$2.5\%\ \\
           & {\footnotesize $S/N([3.6])<2$] $\wedge $[$UBVizYJ>27$]}                                                          &    &                      &        & \\

\enddata
\label{tab:contam}
\tablecomments{
The above selection criteria are outlined explicitly in
\citet{bouwens09a}, for optical-only selection, and
\citet{bouwens14a}, for optical and near-IR selection, yet the
selection method is very similar to other high-$z$ dropout selection
techniques outlined in \citet{bunker03a}, \citet{giavalisco04b},
\citet{beckwith06a}, \citet{stanway07a}, \citet{oesch10a},
and \citet{bouwens11a}. 
Magnitudes are interpolated from the observed COSMOS broad-,
intermediate- and narrow-band imaging to the given selection filters
and the effective area of this search is performed in the
1.76\,deg$^{2}$ area of UltraVISTA in COSMOS ($\simgt$6 times the area
used in the deep dropout searches).
In addition to the stated color selection criteria, we use a S/N
cutoff in $UBVizYJ$ bands as appropriate.
The characteristics of the DSFGs satisfying the selection criteria are
given: their average apparent magnitudes, implied $\beta$ if at the
target redshift, their median photometric redshift \citep[from the
  UltraVISTA near-IR selected catalog;][]{mccracken12a} and their
density on the sky.
The last column gives the percentage of LBG candidates, selected at
the given target redshift, that are in fact DSFGs at lower redshifts.
We use the stated number of candidates identified per square arcminute
as stated in \citet{bouwens09a} and \citet{bouwens14a}.
}
\end{deluxetable*}

Two plausible explanations for this observed `reddening' of
matched-$L_{\rm IR}$ galaxies seen between $0.6<z<1.4$ are (a) an
increasing metallicity of galaxies towards lower redshifts, or (b)
different star formation histories present in galaxies at $z=1.4$
vs.\ $z=0.6$.  In the former case, it should be noted that some studies
investigating the rest-frame UV continuum properties of
optically-selected galaxy populations show evidence for intrinsically
bluer colors in lower metallicity systems
\citep{alavi14a,castellano14a}, perhaps owing to a more top-heavy IMF
in lower metallicity galaxies \citep[e.g.][]{marks12a}.  Note also
that \citet{marks12a} argue that a more top-heavy IMF will manifest
in environments with dense molecular clouds like starbursts, resulting
in bluer rest-frame UV slopes.

The second explanation for the observed reddening with epoch is a
different star formation history between LIRGs at $z=0.6$ and $z=1.4$.
Following \citet{kong04a}, we note
that galaxies with more recent burst histories and younger stellar
populations are expected to have bluer rest-frame UV slopes.  This
suggests, in fact, that our $z=1.4$ DSFGs are likely {\it younger} and
{\it burstier} than their $z=0.6$ analogues, a notion which might
seem contradictory to recent works suggesting that LIRGs and ULIRGs at
high-$z$ are {\it less} likely to be burst-driven than their local
counterparts, based on the galaxy main-sequence 
\citep[e.g.][]{noeske07a,elbaz11a,rodighiero11a,nordon12a}.

Besides redshift evolution, the large scatter in color in each
luminosity bin could also be partly due to differences in the
intrinsic dust attenuation curves, where flatter curves indicate
larger attenuations (\irx) for the same color
\citep[$\beta$][]{gordon00a,burgarella05a}, or viewing angle effects.
Studying the morphological characteristics of the bluest and reddest
sources of a given luminosity or star formation rate to probe
inclination, bulge-to-disk ratio, and interactions will be a necessary
and worthy follow-up study yet is beyond the scope of this paper.

It is worth highlighting that, although we offer many plausible
physical explanations here to describe the deviation (or possible
bluewards $z$-evolution), we primarily attribute the bulk shift off
\irxb\ to dust geometry effects. As we will later discuss in more
detail in \S~\ref{sec:discussion}, we know that nearby LIRGs and
ULIRGs are primarily enshrouded in a thick cocoon of dust, where the
IR emission is high and the UV photons are few.  The UV light that
does escape does so in a patchy pattern, leaking out in bright
concentrations.  These small openings dominate the UV luminosity,
therefore the total UV color of the galaxy, while the dust enshrowded
component (which is spatially dis-associated) dominates $L_{\rm IR}$.
We return to this in our discussion.

Note that the various other measurements of the \irxb\ relation at
$z>0$ from LBGs and DOGs, align with the observed deviation we measure
for DSFGs at their respective IR luminosities, within uncertainty.
Could this explain some of the discrepancies between, e.g. the
\citet{reddy12a} result$-$LBGs that lie blueward of our nominal
\irxb\ fit$-$the \citet{heinis13a} results$-$UV selected galaxies
lying very close to our nominal \irxb\ fit$-$and the \citet{penner12a}
results$-$where dust-selected galaxies are notably bluer than our
nominal \irxb\ fit?  We think yes, that it may be understood as a
function of these' galaxies total star formation rates or IR
luminosities.  So while earlier we thought it auspicious that the
$z\sim2$ \citeauthor{reddy12a} LBGs aligned perfectly with the
\citet{meurer99a} \irxb\ relation (a relation with known biases and
measurement discrepancies), here we can attribute that alignment to
two different biases which together are skewed blueward of the nominal
\irxb\ relation for a heterogeneous, normal galaxy population.  It is
also not surprising to see the much bluer colors in the
\citet{penner12a} DOGs given their high IR luminosities.

\section{How do DSFGs impact high-$z$ galaxy searches?}\label{sec:highz}

The search for the highest redshift galaxies, at $z>4$, is predicated
on the Lyman Break dropout technique \citep{steidel96a}.  DSFGs are
usually thought to be too intrinsically red and too rare to impact
high-$z$ LBG searches, but our results hint that a subset of DSFGs
might satisfy LBG selection criteria, either because they are bluer
than anticipated or, alternatively, their rest-frame optical emission
lines might contribute substantially to broad-band photometry and be
mistaken for a Lyman break at higher redshift
\citep[cf.][]{bouwens11b}.  Although the LBG dropout selection
technique has some advantages, in providing an easily repeatable
selection, it is potentially prone to more contamination than high-$z$
photometric redshift techniques, which are less biased against LBGs
with intrinsically redder colors
\citep{finkelstein10a,finkelstein12a}.  This section makes use of the
extensive, deep optical and near-IR data in the COSMOS field to
explore the extent to which directly-detected \herschel\ DSFGs might
contaminate high-$z$ LBG `dropout' searches.

LBG `dropout' selection at high-$z$ is defined usually in terms of
{\it Hubble Space Telescope} broad band filters.
We use the 30+ band photometry in the COSMOS field \citep[most
  recently complete with deep near-IR imaging from UltraVISTA covering
  1.76\,deg$^2$;][]{mccracken12a} to interpolate and predict
magnitudes in these {\it HST} filters and then apply the various LBG
selection criteria to our DSFGs, as outlined in
Table~\ref{tab:contam}.  We use the LBG dropout selections that are
outlined in \citet{bouwens09a}, for optical data only, and
\citet{bouwens14a}, for optical and near-infrared data.  In addition
to the color cuts, we apply a magnitude cutoff of 27 [AB] in bands
shortward of the \lya\ break.  This prevents against obvious, lower
redshift bright contaminants and is preferred to a strict S/N cutoff
since the COSMOS data are of varying depths compared to the deep {\it
  HST} imaging used to identify $z\simgt4$ LBGs.  The contaminating
fraction of DSFGs in LBG searches is then computed by comparing their
surface densities to those of LBG candidates of comparable magnitudes
\citep[as presented in][]{bouwens09a,bouwens14a}.

Table~\ref{tab:contam} summarizes our results.  We find that at low
redshifts, $z\simlt4$, the contamination of LBG searches with DSFGs is
negligible ($\ll1\%$), since LBGs at that epoch are extremely common
relative to DSFGs.  The DSFGs which do satisfy the LBG color cuts at
these epochs do seem to sit at the appropriate redshifts and are the
bluest of DSFGs, but again, are negligible in number when compared to
the much more common LBG.
Without near-infrared selection criteria, dusty galaxies can easily
pass LBG selection at $z\sim5-6$ and comprise a significant fraction
of contaminants, as much as $\approx$9\%\ at $z\sim6$.  However,
thanks to the recent deep surveys from WFC3 in the near-infrared, most
of those contaminants are thrown out, with rates as low as 0.1\%.

\begin{figure}
\centering
\includegraphics[width=0.99\columnwidth]{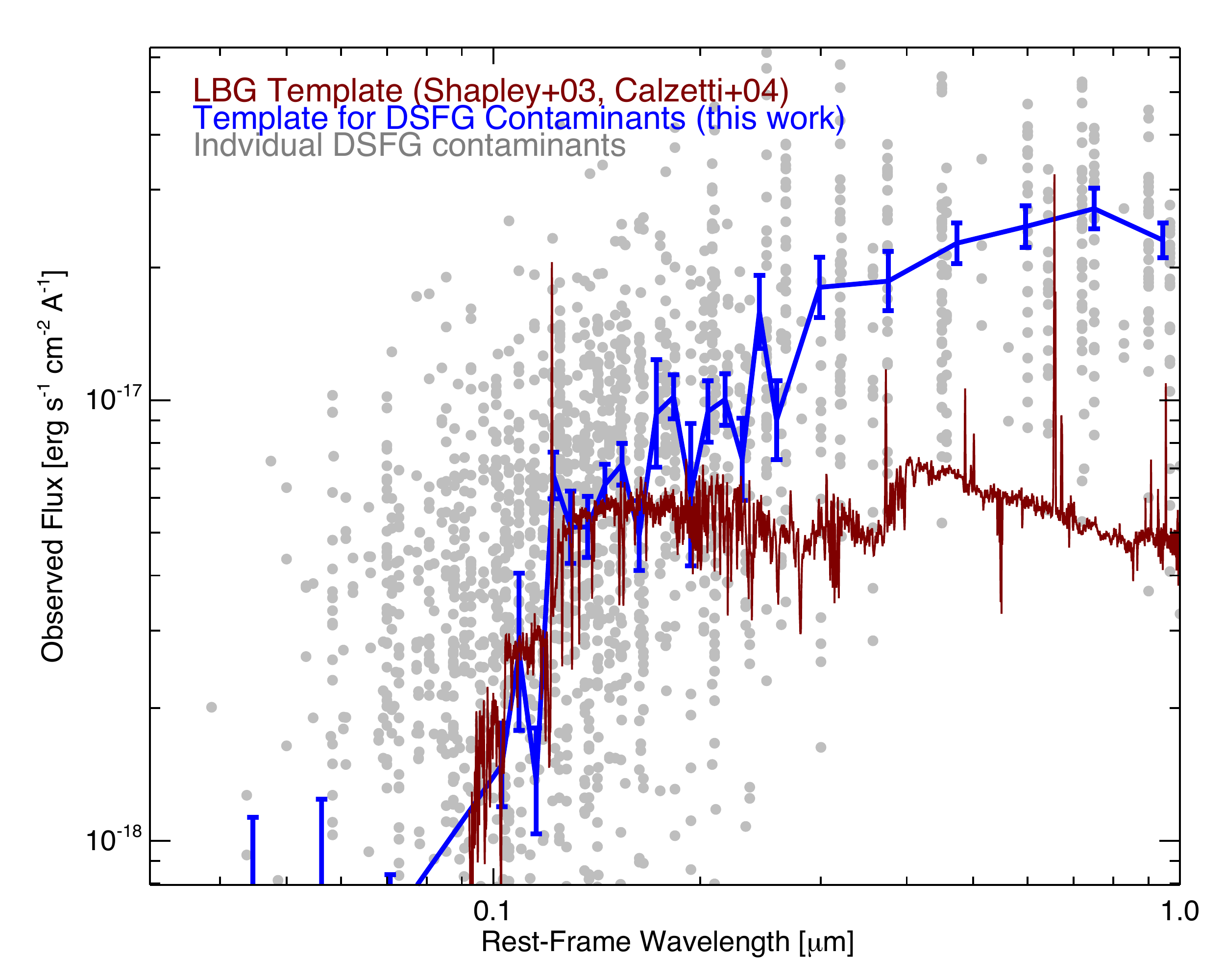}
\caption{Average template SED for the DSFG population that
  contaminates LBG dropout searches, given in rest-frame wavelength at
  the intended target redshift.  In gray, we show the individual
  photometric points for DSFG contaminants which satisfy each of the
  different redshift LBG cuts given in Table~\ref{tab:contam}.  The
  blue curve gives the average contaminant SED for the entire DSFG
  contaminating population, averaged over all selection redshifts (in
  most cases, the DSFG redshifts are dissimilar to the target
  redshifts).  This contaminant template illustrates the need for
  rest-frame optical constraints on LBG SEDs that are anticipated
  to be slightly bluer than most DSFG contaminants.  Note that very
  deep coverage shortward of the Lyman break does not guarantee
  high-redshift identification, since most DSFGs also drop out of the
  bluest optical filters.  The DSFGs' photometry is observed and no
  normalization to a flux scale is done.}
\label{fig:highzcontam}
\end{figure}

Figure~\ref{fig:highzcontam} shows an average template for DSFGs that
contaminate these high-$z$ LBG dropout searches.  To construct the
template, we first build a median template for contaminants in each
LBG selection, corresponding to different redshifts, and then shift
them into the same rest-frame wavelength grid: that which corresponds
to the assumed redshift of selected LBGs.  The median template for all
of the LBG selection is then shown in blue\footnote{Available for
  download now at herschel.uci.edu/cmcasey/research.html under `Tools'
  and in the future available on the PI's research website under the
  same header.}, clearly showing the corresponding break that is
mistaken for the Lyman-$\alpha$ break. Note that this template is not
meant to be physical as it is a coaddition of DSFGs' observed SEDs at
a variety of redshifts, and as such, the break observed is caused by a
number of different physical processes, e.g. bright [O{\sc ii}] emission,
bright \ha, or a 4000\,\AA\ or Balmer break.  One primary difference
seen between the expected SED of an LBG \citep[see the composite in
  Fig~\ref{fig:highzcontam}][]{shapley03a,calzetti94a} and our
contaminant template is that the rest-frame optical emission of LBGs
is bluer than in DSFG contaminants.

Although this method of checking for dusty `contaminants' in UV galaxy
searches has caveats, most notably mismatch photometric depths of the
surveys and the likely incompleteness of DSFG samples beyond $z>2$, it
serves as a worthy reality check for high-$z$ galaxy search campaigns.
The increasing rate of contamination in the highest redshift bins is
due to both the degeneracies of colors over a relatively short
rest-frame wavelength span and the diminishing surface density of
higher-$z$ LBGs.  Overall, contamination rates of $<2-3\%$ from dusty
starbursts bode well for high-$z$ searches.

Besides providing an essential measurement of the Universe's star
formation rate density at very early times, high-$z$ LBG searches have
led us to infer the dust content of the earliest galaxies using the
\irxb\ relation \citep{bouwens09a}.  Because the distribution of
rest-frame UV slopes of LBGs is bluer at earlier times,
\citeauthor{bouwens09a} argued that these high-$z$ LBGs contain very
little dust. Despite the degeneracy between blue dust-less and dusty
galaxies, our results$-$low contamination rates from DSFGs in LBG
searches$-$corroborate the \citeauthor{bouwens09a} conclusions, that
the majority of high-$z$ LBGs contain relatively little dust, by
demonstrating that very few high-$z$ LBGs will be directly detected
with {\it Herschel} or similar submm instruments.  Of course deep
submillimeter follow-up (e.g. from ALMA) is needed to infer the dust
content in individual high-$z$ LBGs \citep[e.g.][]{chapman09b}, in
particular to constrain their SFRs and contribution to the total star
formation rate density.  This is because our results have only ruled
out the tip of the iceberg: it is quite possible that blue LBGs in the
early Universe are much dustier than their UV slopes may infer yet
still too faint to be detected in existing \herschel\ data.

Furthermore, the existence of extreme, dusty galaxies at the same
epochs should raise concern for our understanding of the integrated
star formation rate density \citep[SFRD, e.g.][]{hopkins06a} at these
early times.  The star formation rate in one DSFG can often exceed the
SFR in individual LBGs by factors of 50--100.  While here we have
determined that a very low fraction of LBGs will be directly-detected
by {\it Herschel}, we note that: (a) {\it Herschel} is far less
sensitive to detecting $z>2$ starbursts than longer-wavelength submm
surveys \citep*[at $\approx$1\,mm, see figure 7 of][]{casey14a}; and
(b) very few $z\sim4-6$ DSFGs will be selected as LBGs.  Of the 39
COSMOS DSFGs with photometric redshifts above 4, only 20\%\ satisfy LBG
selection criteria.  Furthermore, we know that high-$z$ extreme DSFGs
like HFLS3 at $z=6.34$ {\it do} exist \citep{riechers13a,cooray14a},
yet the difficulty in obtaining spectroscopic identification has made
the assessment of their volume density quite challenging
\citep{dowell14a}.  While our comparison to LBGs illustrates that
there is little concern for significant contamination, we emphasize
that the $z>4$ SFRD is still highly uncertain without concrete
constraints from direct far-infrared measurements.

\section{Discussion}\label{sec:discussion}

Our results indicate that galaxies with particularly high star
formation rates, most of which are measured from their output in the
far-infrared, are bluer than expected given the nominal dust attenuation
curve for low-luminosity, `normal' star-forming galaxies.  What
underlying physical processes could be responsible for these bluer UV
colors?

The first possible explanation is that the rest-frame UV emission and
far-infrared emission are physically dis-associated, or spatially
distinct \citep{goldader02a,chapman04b}.  Assuming that the systems
are still physically bound, this might be due to a recent catastrophic
event, like a merger-driven burst.  Can geometric effects alone
explain the systematically bluer rest-frame UV slopes in DSFGs?  If a
system consists of two dominant components, one with $L_{\rm
  UV}\approx L_{\rm IR}$ (unobscured) and one with $L_{\rm IR}\gg
L_{\rm UV}$ (obscured), both of which follow the \irxb\ relation, the
\irx\ value assumed for the whole system will be taken from the
obscured component, since there the IR luminosity dominates, and the
$\beta$ value from the unobscured component, which dominates all UV
light; the result is an unresolved, blue DSFG with high \irx. But take
another system, which has its IR luminosity and UV luminosity well
distributed in, e.g., a disk (for a simple test, if you were to split
it into two components, as before, both would have roughly evenly
matched $L_{\rm IR}$ and $L_{\rm UV}$).  If the individual regions of
this homogenized system all followed \irxb, then the total would as
well.  Only dramatic spatial variations in the distribution of $L_{\rm UV}$
and $L_{\rm IR}$ can lift a system substantially above \irxb.

\begin{figure}
\includegraphics[width=0.99\columnwidth]{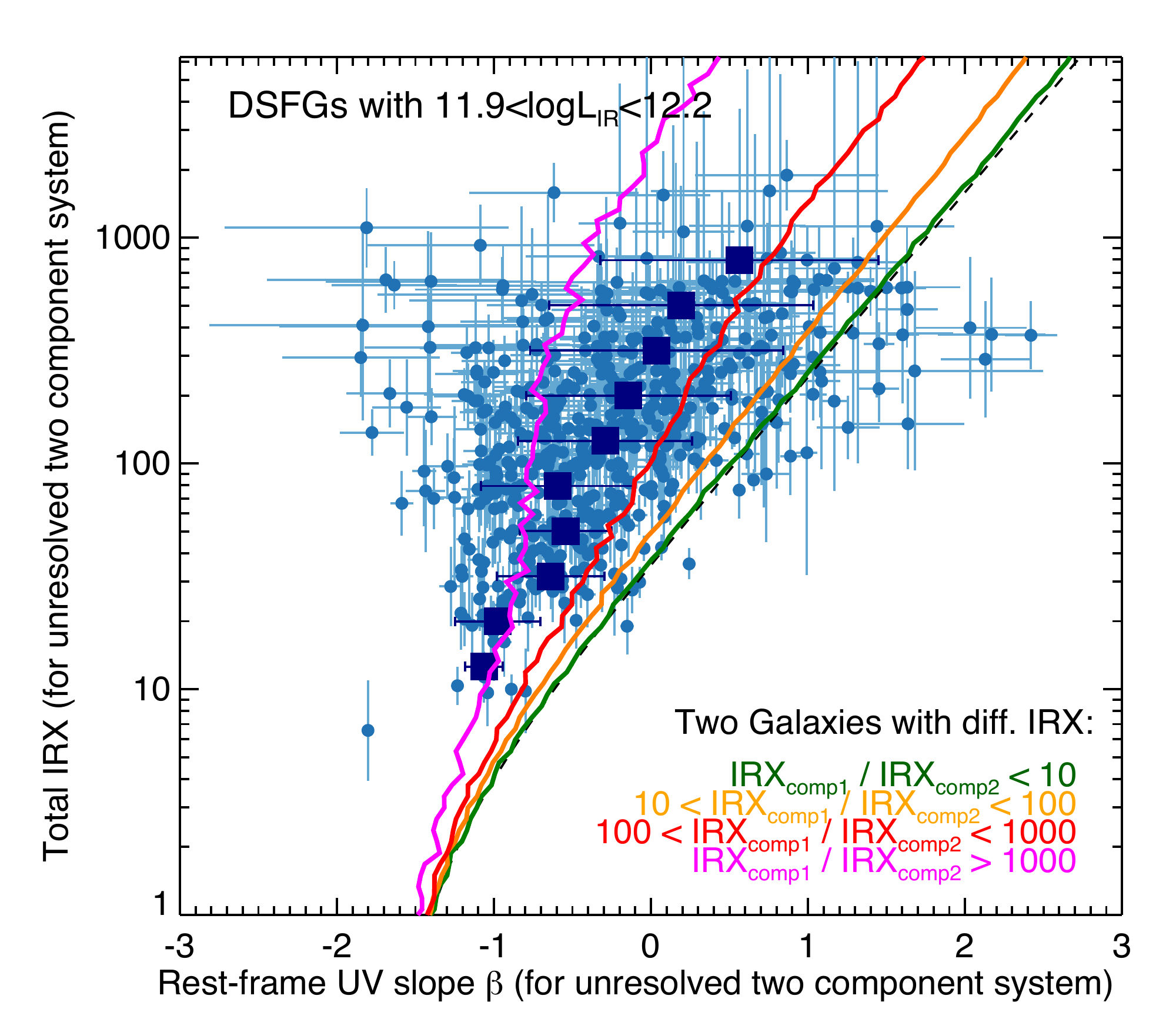}
\caption{Can spatially disassociated UV- and IR-bright components
  explain the bluer rest-frame UV colors of DSFGs?  Here we explore
  how two nearby galaxy components of different obscurations
  (\irx\ values) might be perceived on the \irxb\ plane if the two
  components are not resolved from one another.  We simulate 10$^6$
  hypothetical pairs where all individual components lie on the
  \irxb\ relation (dashed black line) with observed local scatter
  $\pm$0.59 in $\beta$.  Only pairs with significant differences
  between component \irx\ values will be measured as having
  bluer-than-expected colors due to spatial disassociation.  This
  simplified simulation is effectively modeling the relative
  bimodality of a galaxy's dust obscuration.
  We overplot all DSFGs within the narrow $L_{\rm IR}$ ranging
  corresponding to our simulation in light blue (median values are
  dark blue squares) for comparison.
This figure suggests that most 10$^{12}$\lsun\ DSFGs are bimodal, or
have spatially disassociated components which differ in \irx\ by
factors of $\sim$300.  }
\label{fig:geometry}
\end{figure}

Figure~\ref{fig:geometry} illustrates how the relative dustiness
(\irx\ values) between the two hypothetical, nearby components impacts
how `blue' the aggregate sum of the two will be relative to \irxb.  We
constructed this figure by simulating $10^6$ pairs of components,
where each indidivdual component sits on the \irxb\ relation.  To
explain a sample of very blue DSFGs systematically offset from
\irxb\ using this geometrical argument, the vast majority of
DSFGs would need to be comprised of two radially different components
(where \irx\ of the first would be a factor of $\simgt$100--1000 times
larger than the second).  In other words, the distributions of
\irx\ and $\beta$ over the spatial extent of a DSFG would need to be
bimodal. Indeed, this is effectively what we see locally
(e.g. Arp\,220, Mrk\,273, IRAS\,19254$-$7245). The subcomponents of
local ULIRGs which are infrared-bright produce very little UV light
and the patches which are UV-bright are blue and see
little IR contribution.

Aside from physical dis-association, alternate explanations also
exist, although with less observational backing amongst local analogs.
Nevertheless, we discuss them here since the effect we observe could
be due in part to these multiple factors.  As described in
\S~\ref{sec:analysis}, \citet{kong04a} suggest that \irxb\ should not,
in fact, be a fixed relation for all galaxies, but rather it should
vary for galaxies with different star formation histories.  The more
recent and intense the star-formation, the bluer the galaxy (for a
fixed \iruv\ ratio).  While overall this interpretation explains
Figure~\ref{fig:dbeta} qualitatively, we note that such a pronounced
`break' above $L_{\rm IR}\approx10^{11.5}$\lsun\ might not be
expected.  Whether or not it should be expected depends significantly
on the given model assumptions for the effective absorption curve, and
whether or not stars irradiate just their birth clouds or the ambient
ISM \citep*[suggested to transition at a stellar age of
  10\,Myr;][]{charlot00a}, as well as the distribution of the stars
with respect to the ISM, the lifetime of burst episodes (as short as
10\,Myr to as long as 300\,Myr), and the characteristics of the
underlying long-term star formation.  \citet{kong04a} provide a
sensible model construct for quiescent and normal galaxies, but do not
offer an explanation for sources at low $\beta$ and high \irx.  Those
galaxies could be caught during a very short duration burst, where a
substantial fraction of the emitted UV radiation is trapped in stars'
birth clouds before they have time to migrate out.

In addition to having a much bluer intrinsic UV slope, DSFGs might
have enhanced \iruv\ ratios due to the short timescale of recent star
formation.  This might be due to UV radiation being trapped in stars'
birth clouds, or more globally, could be explained by the geometry of
dust and stars in the galaxy.  In a well-mixed ISM with a more
spheroidal distribution, the surface-to-volume ratio is lower, and a
galaxy with higher SFR will have a higher \iruv\ ratio, assuming
$L_{\rm IR}$ emanates from the whole volume while $L_{\rm UV}$ emerges
only from regions close to the surface.  On the contrary, lower
luminosity, continuous-SFR disk galaxies have lower optical depth for
dust attenuation and greater surface area-to-volume ratios, implying
(a) $L_{\rm IR}$ and $L_{\rm UV}$ emanate from the whole volume, and
(b) the underlying UV continuum is redder than in extreme burst
galaxies.

Whether or not the magnitude of the shift towards bluer $\beta$ (or
higher \irx) for DSFGs is in line with expectation requires more
detailed model investigation, which is beyond the scope of this paper.
Nevertheless, our results support the notion that DSFGs are likely
inconsistent with prolonged, constant star formation histories of disk
galaxies, as might be suggested by works favoring the galaxy
main-sequence view of DSFGs at high-$z$.

In the context of the galaxy main-sequence, readers might be curious
how our results would present themselves if, instead of a $L_{\rm
  IR}$- or SFR-dependent break in $\Delta\beta$, we investigated the
break as a function of galaxy stellar mass or specific star formation
rate.  We have intentionally avoided the use of stellar mass estimates
in this paper, primarily due to their reliance on an assumed star
formation history.  Infrared-luminous galaxies' star formation
histories are particularly difficult to determine and have large
uncertainties \citep{hainline11a,michaowski12a,michaowski14a}.  This
in itself is a topic of ongoing debate in the literature \citep*[see
  more in \S~5 of][]{casey14a}.  
We hope that follow-up studies that do investigate \irxb\ with stellar
mass will carefully disentangle various star formation history {\it a
  priori} assumptions from any conclusions used to interpret galaxy
evolution.

Delving deeper into the underlying physics behind bluer UV slopes,
aside from our simple and favored geometric explanation, the
possibility was raised in \S~\ref{sec:analysis} that metallicity and a
possible top-heavy IMFs might also contribute to a higher proportion
of O starlight, and thus bluer slopes.  A top-heavy IMF might be
plausible in the dense star-forming regions in ultraluminous galaxies,
e.g. DSFGs, consistent with dense star clusters in the Milky Way like
Arches and Westerlund~1 \citep{marks12a}.  Furthermore, our
observation of some evolution in DSFG UV slopes of matched IR
luminosity in the range $0.6<z<1.4$ would be consistent with the idea
that higher redshift systems are lower metallicity systems.  Note
that, if we assume that DSFGs at higher redshifts should be less
bursty \citep[as many papers suggest despite direct evidence to the
  contrary, e.g.][]{engel10a,ivison12a}, we might expect a shift in
the {\it opposite} direction, towards redder UV slopes, or no shift at
all.  This is because main sequence galaxies are proposed to have
steady state, approximately constant star formation rates for most of
their lifetime; even at very high SFRs ($\simgt$100\sfr), the expected
underlying (unextincted) UV continuum of steady state star formation
will be redder than starburst galaxies, which have a
disproportionately large contribution of light from O and B stars
above the less luminous and older A stars.  Again, our results are
inconsistent with the suggestion that $>10^{11.5}$\lsun\ DSFGs at
high-$z$ are dominated by steady state, secular, disky star formation.

\section{Conclusions}\label{sec:conclusions}

This paper has investigated the relationship between the \iruv\ ratio
(probing the relative `dustiness' of galaxies) to their rest-frame
ultraviolet continuum slope, $\beta$, particularly as it pertains to
infrared-selected galaxies.  By comparing a sample of $\approx$1200
nearby galaxies ($z<0.085$) spanning star formation rates
0.03--300\sfr\ to a large sample of $\approx$4000
IR-selected galaxies spanning photometric redshifts $0<z<5$ in the
COSMOS field, we have arrived at the following conclusions.
\begin{enumerate}
\item We derive a much redder \irxb\ relationship for local galaxies
  than was presented in some of the original works on the topic
  \citep[e.g.][]{meurer99a} and attribute the difference to (a)
  differences between {\it IUE} and \galex\ aperture limitations (the
  latter provides a more representative characteristic estimate for
  $\beta$); and (b) the fact that we derive the relation for a more
  heterogeneous population, not limited to galaxies selected as blue,
  compact starbursts.  Our derived relation is given in
  Equation~\ref{eq:irxbeta}.  This is roughly consistent with prior
  works, which corrected the original relation for aperture
  differences \citep{takeuchi12a}.
\item We find that, at both low and high-redshift, DSFGs with high
  SFRs above $\approx50$\sfr deviate from this \irxb\ relation towards
  bluer colors, where the offset grows with increasing IR luminosity
  and star formation rate.  The deviation towards bluer colors,
  measured as $\Delta\beta$ is seen both in local and high-$z$ samples
  above a `break' IR luminosity of $\approx10^{11-11.5}$\lsun\ and
  increases with increasing luminosity such that at $L_{\rm
    IR}=10^{13}$\lsun, galaxies are on average $\Delta\beta=-2$ bluer
  than expected from the nominal \irxb\ relation fit in
  Eq~\ref{eq:irxbeta}.  This offset towards bluer colors is shown not
  to be caused by sample selection effects.
\item Subtle redshift evolution is detected in the narrow luminosity
  regime $4-7.5\times10^{11}$\lsun\ for $0.6<z<1.4$, where galaxies
  of matched IR luminosity are on average $\Delta\beta=-1$ bluer at
  $z=1.4$ than at $z=0.6$.  No redshift evolution is detected above
  $z>1.4$ (at representatively higher IR luminosities, corresponding to
  detection limits at that epoch).  More extensive samples of equal
  luminosity over a wider range of epochs is needed to verify this
  perceived evolution.
\item We attribute the bluer colors in dusty galaxies to more recent,
  rapid episodes of star formation \citep[following the `birthrate'
    parameter model offered by][]{kong04a}, where more IR luminous
  galaxies have a more prominent population of young, O and B stars
  (contributing to the rest-frame far-UV emission) than galaxies of
  more modest star formation rates.  Not only are they intrinsically
  bluer, DSFGs likely have higher \irx\ values caused by lower
  emergent UV-to-IR luminosity ratios and mixed, patchy geometry.
  This is consistent with the idea that star formation in DSFGs at
  high redshift is dominated by burst activity rather than steady
  state, gradual disk growth.
\item With deep multi-band UV/optical and near-IR data in hand, we
  investigated the rates at which DSFGs contaminate high-$z$ LBG
  dropout searches.  Due to the relatively low sky density of DSFGs,
  we find very low contamination rates at $z\simlt7$ when both deep
  optical and near-infrared data exist.  Contamination rates increase
  at higher redshifts, up to 5.1\%\ at $z\sim10$, where there is less
  information on the rest-frame optical and more potential
  low-redshift interlopers (likewise, we see high contamination,
  8.9\%, at $z\sim6$ if observed near-infrared bands are not used for
  LBG selection).  Overall, DSFG contamination rates $<10$\%\ bode
  well for LBG searches; however, we caution that this does not imply
  that $z>4$ SFRD estimates from LBGs alone are sufficient for
  understanding star formation at early times.
\end{enumerate}
While this work has shed light on the issue of rest-frame ultraviolet
emission in the dustiest galaxies, it is clear that much more work is
necessary to understand the underlying physics driving these bluer
rest-frame UV slopes in dusty galaxies.  Geometry, morphological
effects, and galaxy interactions need to be constrained on individual
systems.  Constraining these galaxies' star formation histories via
detailed SED fitting is crucial to isolate the dominant physical
mechanisms in high-$z$ star formation.  The coming years will provide
crucial insight into the issue of extinction in extreme galaxies, by
studying its links to gas dynamics, star formation rate, galaxy
morphology, gas supply, dust reservoir, metallicity, and differences
between emission line and continuum extinction.  This paper has only
provided a broad context from which necessary follow-up studies of
detailed systems will reveal the nature of dust attenuation in extreme
environments.

\acknowledgements

COSMOS is based on observations with the NASA/ESA {\it Hubble Space
  Telescope}, obtained at the Space Telescope Science Institute, which
is operated by AURA Inc, under NASA contract NAS 5-26555; also based
on data collected at: the Subaru Telescope, which is operated by the
National Astronomical Observatory of Japan; the XMM-Newton, an ESA
science mission with instruments and contributions directly funded by
ESA Member States and NASA; the European Southern Observatory, Chile;
Kitt Peak National Observatory, Cerro Tololo Inter-American
Observatory, and the National Optical Astronomy Observatory, which are
operated by the Association of Universities for Research in Astronomy,
Inc. (AURA) under cooperative agreement with the National Science
Foundation; the National Radio Astronomy Observatory which is a
facility of the National Science Foundation operated under cooperative
agreement by Associated Universities, Inc; and the
Canada-France-Hawaii Telescope operated by the National Research
Council of Canada, the Centre National de la Recherche Scientifique de
France and the University of Hawaii.

PACS has been developed by a consortium of institutes led by MPE
(Germany) and including UVIE (Austria); KU Leuven, CSL, IMEC
(Belgium); CEA, LAM (France); MPIA (Germany); INAF-IFSI/OAA/OAP/OAT,
LENS, SISSA (Italy); IAC (Spain).  This development has been supported
by the funding agencies BMVIT (Austria), ESA-PRODEX (Belgium),
CEA/CNES (France), DLR (Germany), ASI/INAF (Italy), and CICYT/MCYT
(Spain).

SPIRE has been developed by a consortium of institutes led by Cardiff
Univ. (UK) and including: Univ. Lethbridge (Canada); NAOC (China);
CEA, LAM (France); IFSI, Univ. Padua (Italy); IAC (Spain); Stockholm
Observatory (Sweden); Imperial College London, RAL, UCL-MSSL, UKATC,
Univ. Sussex (UK); and Caltech, JPL, HNSC, Univ. Colorado (USA).  This
development has been supported by national funding agencies: CSA
(Canada); NAOC (China); CEA, CNES, CNRS (France); ASI (Italy); MCINN
(Spain); SNSB (Sweden); STFC, UKSA (UK); and NASA (USA).

This research has made use of data from the HerMES project
(http://hermes.sussex.ac.uk/).  HerMES is a \herschel\ Key Programme
utilizing Guaranteed Time from the SPIRE instrument team, ESAC
scientists and a mission scientist.  The data presented in this paper
will be released through the HerMES Database in Marseille, HeDaM
(http://hedam.oamp.fr/HerMES/).  We would also like to recognize the
use of the Glue Visualization tool (www.glueviz.org) in the initial
analysis of this data-set.

CMC acknowledges support from a McCue Fellowship through the
University of California, Irvine's Center for Cosmology.  AC
acknowledge support from NSF AST-1313319.  RJI acknowledges support
from ERC AdG, COSMICISM.  TTT has been supported by the Grant-in-Aid
for the Scientific Research Fund (24111707) commissioned by the
Ministry of Education, Culture, Sports, Science and Technology (MEXT)
of Japan. TTT has also been partially supported from Strategic Young
Researches Overseas Visits Program for Accelerating Brain Circulation
from the MEXT.  We are grateful to the anonymous referee for
productive comments which improved the paper. We would also like to
thank Naveen Reddy, Brian Siana, Mara Salvato and Douglas Scott for
helpful conversations during the manuscript's preparation.


\begin{thebibliography}{142}
\expandafter\ifx\csname natexlab\endcsname\relax\def\natexlab#1{#1}\fi

\bibitem[{{Adelberger} \& {Steidel}(2000)}]{adelberger00a}
{Adelberger}, K.~L. \& {Steidel}, C.~C. 2000, \apj, 544, 218

\bibitem[{{Alavi} {et~al.}(2014)}]{alavi14a}
{Alavi}, A. {et~al.} 2014, \apj, 780, 143

\bibitem[{{Aretxaga} {et~al.}(2011)}]{aretxaga11a}
{Aretxaga}, I. {et~al.} 2011, \mnras, 415, 3831

\bibitem[{{Armus} {et~al.}(2009)}]{armus09a}
{Armus}, L. {et~al.} 2009, \pasp, 121, 559

\bibitem[{{Bauer} {et~al.}(2011){Bauer}, {Conselice}, {P{\'e}rez-Gonz{\'a}lez},
  {Gr{\"u}tzbauch}, {Bluck}, {Buitrago}, \& {Mortlock}}]{bauer11a}
{Bauer}, A.~E., {Conselice}, C.~J., {P{\'e}rez-Gonz{\'a}lez}, P.~G.,
  {Gr{\"u}tzbauch}, R., {Bluck}, A.~F.~L., {Buitrago}, F., \& {Mortlock}, A.
  2011, \mnras, 417, 289

\bibitem[{{Beckwith} {et~al.}(2006)}]{beckwith06a}
{Beckwith}, S.~V.~W. {et~al.} 2006, \aj, 132, 1729

\bibitem[{{Blain} {et~al.}(2004){Blain}, {Chapman}, {Smail}, \&
  {Ivison}}]{blain04a}
{Blain}, A.~W., {Chapman}, S.~C., {Smail}, I., \& {Ivison}, R. 2004, \apj, 611,
  725

\bibitem[{{Boissier} {et~al.}(2007)}]{boissier07a}
{Boissier}, S. {et~al.} 2007, \apjs, 173, 524

\bibitem[{{Boquien} {et~al.}(2009)}]{boquien09a}
{Boquien}, M. {et~al.} 2009, \apj, 706, 553

\bibitem[{{Boquien} {et~al.}(2012)}]{boquien12a}
---. 2012, \aap, 539, A145

\bibitem[{{Bouwens} {et~al.}(2009)}]{bouwens09a}
{Bouwens}, R.~J. {et~al.} 2009, \apj, 705, 936

\bibitem[{{Bouwens} {et~al.}(2011{\natexlab{a}})}]{bouwens11b}
---. 2011{\natexlab{a}}, \nat, 469, 504

\bibitem[{{Bouwens} {et~al.}(2011{\natexlab{b}})}]{bouwens11a}
---. 2011{\natexlab{b}}, \apj, 737, 90

\bibitem[{{Bouwens} {et~al.}(2014)}]{bouwens14a}
---. 2014, ArXiv e-prints

\bibitem[{{Brusa} {et~al.}(2010)}]{brusa10a}
{Brusa}, M. {et~al.} 2010, \apj, 716, 348

\bibitem[{{Bruzual} \& {Charlot}(2003)}]{bruzual03a}
{Bruzual}, G. \& {Charlot}, S. 2003, \mnras, 344, 1000

\bibitem[{{Buat} {et~al.}(2005)}]{buat05a}
{Buat}, V. {et~al.} 2005, \apjl, 619, L51

\bibitem[{{Buat} {et~al.}(2010)}]{buat10a}
---. 2010, \mnras, 409, L1

\bibitem[{{Buat} {et~al.}(2011)}]{buat11a}
---. 2011, \aap, 533, A93

\bibitem[{{Buat} {et~al.}(2012)}]{buat12a}
---. 2012, \aap, 545, A141

\bibitem[{{Bunker} {et~al.}(2003){Bunker}, {Stanway}, {Ellis}, {McMahon}, \&
  {McCarthy}}]{bunker03a}
{Bunker}, A.~J., {Stanway}, E.~R., {Ellis}, R.~S., {McMahon}, R.~G., \&
  {McCarthy}, P.~J. 2003, \mnras, 342, L47

\bibitem[{{Burgarella} {et~al.}(2005){Burgarella}, {Buat}, \&
  {Iglesias-P{\'a}ramo}}]{burgarella05a}
{Burgarella}, D., {Buat}, V., \& {Iglesias-P{\'a}ramo}, J. 2005, \mnras, 360,
  1413

\bibitem[{{Calzetti}(2001)}]{calzetti01a}
{Calzetti}, D. 2001, \pasp, 113, 1449

\bibitem[{{Calzetti} {et~al.}(1994){Calzetti}, {Kinney}, \&
  {Storchi-Bergmann}}]{calzetti94a}
{Calzetti}, D., {Kinney}, A.~L., \& {Storchi-Bergmann}, T. 1994, \apj, 429, 582

\bibitem[{{Capak} {et~al.}(2007)}]{capak07a}
{Capak}, P. {et~al.} 2007, \apjs, 172, 99

\bibitem[{{Caputi} {et~al.}(2007){Caputi}, {Lagache}, {Yan}, {Dole},
  {Bavouzet}, {Le Floc'h}, {Choi}, {Helou}, \& {Reddy}}]{caputi07a}
{Caputi}, K.~I., {Lagache}, G., {Yan}, L., {Dole}, H., {Bavouzet}, N., {Le
  Floc'h}, E., {Choi}, P.~I., {Helou}, G., \& {Reddy}, N. 2007, \apj, 660, 97

\bibitem[{{Cardelli} {et~al.}(1989){Cardelli}, {Clayton}, \&
  {Mathis}}]{cardelli89a}
{Cardelli}, J.~A., {Clayton}, G.~C., \& {Mathis}, J.~S. 1989, \apj, 345, 245

\bibitem[{{Casey}(2012)}]{casey12a}
{Casey}, C.~M. 2012, \mnras, 425, 3094

\bibitem[{{Casey} {et~al.}(2009){Casey}, {Chapman}, {Beswick}, {Biggs},
  {Blain}, {Hainline}, {Ivison}, {Muxlow}, \& {Smail}}]{casey09b}
{Casey}, C.~M., {Chapman}, S.~C., {Beswick}, R.~J., {Biggs}, A.~D., {Blain},
  A.~W., {Hainline}, L.~J., {Ivison}, R.~J., {Muxlow}, T.~W.~B., \& {Smail}, I.
  2009, \mnras, 399, 121

\bibitem[{{Casey} {et~al.}(2014){Casey}, {Narayanan}, \& {Cooray}}]{casey14a}
{Casey}, C.~M., {Narayanan}, D., \& {Cooray}, A. 2014, ArXiv e-prints

\bibitem[{{Casey} {et~al.}(2012{\natexlab{a}})}]{casey12c}
{Casey}, C.~M. {et~al.} 2012{\natexlab{a}}, \apj, 761, 139

\bibitem[{{Casey} {et~al.}(2012{\natexlab{b}})}]{casey12b}
---. 2012{\natexlab{b}}, \apj, 761, 140

\bibitem[{{Casey} {et~al.}(2013)}]{casey13a}
---. 2013, \mnras, 436, 1919

\bibitem[{{Castellano} {et~al.}(2014)}]{castellano14a}
{Castellano}, M. {et~al.} 2014, ArXiv e-prints

\bibitem[{{Chapman} {et~al.}(2005){Chapman}, {Blain}, {Smail}, \&
  {Ivison}}]{chapman05a}
{Chapman}, S.~C., {Blain}, A.~W., {Smail}, I., \& {Ivison}, R.~J. 2005, \apj,
  622, 772

\bibitem[{{Chapman} \& {Casey}(2009)}]{chapman09b}
{Chapman}, S.~C. \& {Casey}, C.~M. 2009, \mnras, 398, 1615

\bibitem[{{Chapman} {et~al.}(2004{\natexlab{a}}){Chapman}, {Smail}, {Blain}, \&
  {Ivison}}]{chapman04a}
{Chapman}, S.~C., {Smail}, I., {Blain}, A.~W., \& {Ivison}, R.~J.
  2004{\natexlab{a}}, \apj, 614, 671

\bibitem[{{Chapman} {et~al.}(2004{\natexlab{b}}){Chapman}, {Smail},
  {Windhorst}, {Muxlow}, \& {Ivison}}]{chapman04b}
{Chapman}, S.~C., {Smail}, I., {Windhorst}, R., {Muxlow}, T., \& {Ivison},
  R.~J. 2004{\natexlab{b}}, \apj, 611, 732

\bibitem[{{Charlot} \& {Fall}(2000)}]{charlot00a}
{Charlot}, S. \& {Fall}, S.~M. 2000, \apj, 539, 718

\bibitem[{{Chary} \& {Elbaz}(2001)}]{chary01a}
{Chary}, R. \& {Elbaz}, D. 2001, \apj, 556, 562

\bibitem[{{Civano} {et~al.}(2012)}]{civano12a}
{Civano}, F. {et~al.} 2012, \apjs, 201, 30

\bibitem[{{Conroy} {et~al.}(2010){Conroy}, {Schiminovich}, \&
  {Blanton}}]{conroy10a}
{Conroy}, C., {Schiminovich}, D., \& {Blanton}, M.~R. 2010, \apj, 718, 184

\bibitem[{{Cooray} {et~al.}(2014)}]{cooray14a}
{Cooray}, A. {et~al.} 2014, ArXiv e-prints

\bibitem[{{Coppin} {et~al.}(2014)}]{coppin14a}
{Coppin}, K.~E.~K. {et~al.} 2014, ArXiv e-prints

\bibitem[{{Cortese} {et~al.}(2006)}]{cortese06a}
{Cortese}, L. {et~al.} 2006, \apj, 637, 242

\bibitem[{{Dale} \& {Helou}(2002)}]{dale02a}
{Dale}, D.~A. \& {Helou}, G. 2002, \apj, 576, 159

\bibitem[{{Dey} {et~al.}(2008)}]{dey08a}
{Dey}, A. {et~al.} 2008, \apj, 677, 943

\bibitem[{{Donley} {et~al.}(2012)}]{donley12a}
{Donley}, J.~L. {et~al.} 2012, \apj, 748, 142

\bibitem[{{Dowell} {et~al.}(2014)}]{dowell14a}
{Dowell}, C.~D. {et~al.} 2014, \apj, 780, 75

\bibitem[{{Dunne} {et~al.}(2003){Dunne}, {Eales}, {Ivison}, {Morgan}, \&
  {Edmunds}}]{dunne03a}
{Dunne}, L., {Eales}, S., {Ivison}, R., {Morgan}, H., \& {Edmunds}, M. 2003,
  \nat, 424, 285

\bibitem[{{Elbaz} {et~al.}(2011)}]{elbaz11a}
{Elbaz}, D. {et~al.} 2011, \aap, 533, A119+

\bibitem[{{Engel} {et~al.}(2010){Engel}, {Tacconi}, {Davies}, {Neri}, {Smail},
  {Chapman}, {Genzel}, {Cox}, {Greve}, {Ivison}, {Blain}, {Bertoldi}, \&
  {Omont}}]{engel10a}
{Engel}, H., {Tacconi}, L.~J., {Davies}, R.~I., {Neri}, R., {Smail}, I.,
  {Chapman}, S.~C., {Genzel}, R., {Cox}, P., {Greve}, T.~R., {Ivison}, R.~J.,
  {Blain}, A., {Bertoldi}, F., \& {Omont}, A. 2010, \apj, 724, 233

\bibitem[{{Finkelstein} {et~al.}(2010){Finkelstein}, {Papovich}, {Giavalisco},
  {Reddy}, {Ferguson}, {Koekemoer}, \& {Dickinson}}]{finkelstein10a}
{Finkelstein}, S.~L., {Papovich}, C., {Giavalisco}, M., {Reddy}, N.~A.,
  {Ferguson}, H.~C., {Koekemoer}, A.~M., \& {Dickinson}, M. 2010, \apj, 719,
  1250

\bibitem[{{Finkelstein} {et~al.}(2012)}]{finkelstein12a}
{Finkelstein}, S.~L. {et~al.} 2012, \apj, 756, 164

\bibitem[{{Frayer} {et~al.}(2000){Frayer}, {Smail}, {Ivison}, \&
  {Scoville}}]{frayer00a}
{Frayer}, D.~T., {Smail}, I., {Ivison}, R.~J., \& {Scoville}, N.~Z. 2000, \aj,
  120, 1668

\bibitem[{{Fu} {et~al.}(2012)}]{fu12a}
{Fu}, H. {et~al.} 2012, \apj, 753, 134

\bibitem[{{Giavalisco} {et~al.}(2004)}]{giavalisco04b}
{Giavalisco}, M. {et~al.} 2004, \apjl, 600, L103

\bibitem[{{Gil de Paz} {et~al.}(2007)}]{gil-de-paz07a}
{Gil de Paz}, A. {et~al.} 2007, \apjs, 173, 185

\bibitem[{{Goldader} {et~al.}(2002){Goldader}, {Meurer}, {Heckman}, {Seibert},
  {Sanders}, {Calzetti}, \& {Steidel}}]{goldader02a}
{Goldader}, J.~D., {Meurer}, G., {Heckman}, T.~M., {Seibert}, M., {Sanders},
  D.~B., {Calzetti}, D., \& {Steidel}, C.~C. 2002, \apj, 568, 651

\bibitem[{{Gordon} {et~al.}(2003){Gordon}, {Clayton}, {Misselt}, {Landolt}, \&
  {Wolff}}]{gordon03a}
{Gordon}, K.~D., {Clayton}, G.~C., {Misselt}, K.~A., {Landolt}, A.~U., \&
  {Wolff}, M.~J. 2003, \apj, 594, 279

\bibitem[{{Gordon} {et~al.}(2000){Gordon}, {Clayton}, {Witt}, \&
  {Misselt}}]{gordon00a}
{Gordon}, K.~D., {Clayton}, G.~C., {Witt}, A.~N., \& {Misselt}, K.~A. 2000,
  \apj, 533, 236

\bibitem[{{Goto} {et~al.}(2010)}]{goto10a}
{Goto}, T. {et~al.} 2010, \aap, 514, A6+

\bibitem[{{Griffin} {et~al.}(2010)}]{griffin10a}
{Griffin}, M.~J. {et~al.} 2010, \aap, 518, L3

\bibitem[{{Hainline} {et~al.}(2011){Hainline}, {Blain}, {Smail}, {Alexander},
  {Armus}, {Chapman}, \& {Ivison}}]{hainline11a}
{Hainline}, L.~J., {Blain}, A.~W., {Smail}, I., {Alexander}, D.~M., {Armus},
  L., {Chapman}, S.~C., \& {Ivison}, R.~J. 2011, \apj, 740, 96

\bibitem[{{Hao} {et~al.}(2011){Hao}, {Kennicutt}, {Johnson}, {Calzetti},
  {Dale}, \& {Moustakas}}]{hao11a}
{Hao}, C.-N., {Kennicutt}, R.~C., {Johnson}, B.~D., {Calzetti}, D., {Dale},
  D.~A., \& {Moustakas}, J. 2011, \apj, 741, 124

\bibitem[{{Hathi} {et~al.}(2008){Hathi}, {Malhotra}, \& {Rhoads}}]{hathi08a}
{Hathi}, N.~P., {Malhotra}, S., \& {Rhoads}, J.~E. 2008, \apj, 673, 686

\bibitem[{{Heinis} {et~al.}(2013)}]{heinis13a}
{Heinis}, S. {et~al.} 2013, \mnras, 429, 1113

\bibitem[{{Hinshaw} {et~al.}(2009)}]{hinshaw09a}
{Hinshaw}, G. {et~al.} 2009, \apjs, 180, 225

\bibitem[{{Hodge} {et~al.}(2012){Hodge}, {Carilli}, {Walter}, {de Blok},
  {Riechers}, {Daddi}, \& {Lentati}}]{hodge12a}
{Hodge}, J.~A., {Carilli}, C.~L., {Walter}, F., {de Blok}, W.~J.~G.,
  {Riechers}, D., {Daddi}, E., \& {Lentati}, L. 2012, \apj, 760, 11

\bibitem[{{Hopkins} \& {Beacom}(2006)}]{hopkins06a}
{Hopkins}, A.~M. \& {Beacom}, J.~F. 2006, \apj, 651, 142

\bibitem[{{Howell} {et~al.}(2010)}]{howell10a}
{Howell}, J.~H. {et~al.} 2010, \apj, 715, 572

\bibitem[{{Ilbert} {et~al.}(2009)}]{ilbert09a}
{Ilbert}, O. {et~al.} 2009, \apj, 690, 1236

\bibitem[{{Ilbert} {et~al.}(2010)}]{ilbert10a}
---. 2010, \apj, 709, 644

\bibitem[{{Ivison} {et~al.}(2010){Ivison}, {Papadopoulos}, {Smail}, {Greve},
  {Thomson}, {Xilouris}, \& {Chapman}}]{ivison10c}
{Ivison}, R.~J., {Papadopoulos}, P.~P., {Smail}, I., {Greve}, T.~R., {Thomson},
  A.~P., {Xilouris}, E.~M., \& {Chapman}, S.~C. 2010, ArXiv e-prints

\bibitem[{{Ivison} {et~al.}(2011){Ivison}, {Papadopoulos}, {Smail}, {Greve},
  {Thomson}, {Xilouris}, \& {Chapman}}]{ivison11a}
---. 2011, \mnras, 412, 1913

\bibitem[{{Ivison} {et~al.}(1998){Ivison}, {Smail}, {Le Borgne}, {Blain},
  {Kneib}, {Bezecourt}, {Kerr}, \& {Davies}}]{ivison98a}
{Ivison}, R.~J., {Smail}, I., {Le Borgne}, J., {Blain}, A.~W., {Kneib}, J.,
  {Bezecourt}, J., {Kerr}, T.~H., \& {Davies}, J.~K. 1998, \mnras, 298, 583

\bibitem[{{Ivison} {et~al.}(2012)}]{ivison12a}
{Ivison}, R.~J. {et~al.} 2012, MNRAS, 425, 1320

\bibitem[{{Kawada} {et~al.}(2007)}]{kawada07a}
{Kawada}, M. {et~al.} 2007, \pasj, 59, 389

\bibitem[{{Kinney} {et~al.}(1993){Kinney}, {Bohlin}, {Calzetti}, {Panagia}, \&
  {Wyse}}]{kinney93a}
{Kinney}, A.~L., {Bohlin}, R.~C., {Calzetti}, D., {Panagia}, N., \& {Wyse},
  R.~F.~G. 1993, \apjs, 86, 5

\bibitem[{{Kong} {et~al.}(2004){Kong}, {Charlot}, {Brinchmann}, \&
  {Fall}}]{kong04a}
{Kong}, X., {Charlot}, S., {Brinchmann}, J., \& {Fall}, S.~M. 2004, \mnras,
  349, 769

\bibitem[{{Kriek} \& {Conroy}(2013)}]{kriek13a}
{Kriek}, M. \& {Conroy}, C. 2013, \apjl, 775, L16

\bibitem[{{Le Floc'h} {et~al.}(2005)}]{le-floch05a}
{Le Floc'h}, E. {et~al.} 2005, \apj, 632, 169

\bibitem[{{Lee} {et~al.}(2012){Lee}, {Alberts}, {Atlee}, {Dey}, {Pope},
  {Jannuzi}, {Reddy}, \& {Brown}}]{lee12a}
{Lee}, K.-S., {Alberts}, S., {Atlee}, D., {Dey}, A., {Pope}, A., {Jannuzi},
  B.~T., {Reddy}, N., \& {Brown}, M.~J.~I. 2012, \apjl, 758, L31

\bibitem[{{Lee} {et~al.}(2013)}]{lee13a}
{Lee}, N. {et~al.} 2013, \apj, 778, 131

\bibitem[{{Lilly} {et~al.}(1995){Lilly}, {Tresse}, {Hammer}, {Crampton}, \& {Le
  Fevre}}]{lilly95a}
{Lilly}, S.~J., {Tresse}, L., {Hammer}, F., {Crampton}, D., \& {Le Fevre}, O.
  1995, \apj, 455, 108

\bibitem[{{Lutz} {et~al.}(2011)}]{lutz11a}
{Lutz}, D. {et~al.} 2011, A\&A, 532

\bibitem[{{Madau} {et~al.}(1996){Madau}, {Ferguson}, {Dickinson}, {Giavalisco},
  {Steidel}, \& {Fruchter}}]{madau96a}
{Madau}, P., {Ferguson}, H.~C., {Dickinson}, M.~E., {Giavalisco}, M.,
  {Steidel}, C.~C., \& {Fruchter}, A. 1996, \mnras, 283, 1388

\bibitem[{{Magdis} {et~al.}(2011)}]{magdis11a}
{Magdis}, G.~E. {et~al.} 2011, \aap, 534, A15

\bibitem[{{Marks} {et~al.}(2012){Marks}, {Kroupa}, {Dabringhausen}, \&
  {Pawlowski}}]{marks12a}
{Marks}, M., {Kroupa}, P., {Dabringhausen}, J., \& {Pawlowski}, M.~S. 2012,
  \mnras, 422, 2246

\bibitem[{{McCracken} {et~al.}(2012)}]{mccracken12a}
{McCracken}, H.~J. {et~al.} 2012, \aap, 544, A156

\bibitem[{{Meurer} {et~al.}(1999){Meurer}, {Heckman}, \&
  {Calzetti}}]{meurer99a}
{Meurer}, G.~R., {Heckman}, T.~M., \& {Calzetti}, D. 1999, \apj, 521, 64

\bibitem[{{Meurer} {et~al.}(1995){Meurer}, {Heckman}, {Leitherer}, {Kinney},
  {Robert}, \& {Garnett}}]{meurer95a}
{Meurer}, G.~R., {Heckman}, T.~M., {Leitherer}, C., {Kinney}, A., {Robert}, C.,
  \& {Garnett}, D.~R. 1995, \aj, 110, 2665

\bibitem[{{Micha{\l}owski} {et~al.}(2012){Micha{\l}owski}, {Dunlop},
  {Cirasuolo}, {Hjorth}, {Hayward}, \& {Watson}}]{michaowski12a}
{Micha{\l}owski}, M.~J., {Dunlop}, J.~S., {Cirasuolo}, M., {Hjorth}, J.,
  {Hayward}, C.~C., \& {Watson}, D. 2012, \aap, 541, A85

\bibitem[{{Micha{\l}owski} {et~al.}(2014){Micha{\l}owski}, {Hayward}, {Dunlop},
  {Bruce}, {Cirasuolo}, {Cullen}, \& {Hernquist}}]{michaowski14a}
{Micha{\l}owski}, M.~J., {Hayward}, C.~C., {Dunlop}, J.~S., {Bruce}, V.~A.,
  {Cirasuolo}, M., {Cullen}, F., \& {Hernquist}, L. 2014, ArXiv e-prints

\bibitem[{{Morrissey} {et~al.}(2007)}]{morrissey07a}
{Morrissey}, P. {et~al.} 2007, \apjs, 173, 682

\bibitem[{{Mu{\~n}oz-Mateos} {et~al.}(2009)}]{munoz-mateos09a}
{Mu{\~n}oz-Mateos}, J.~C. {et~al.} 2009, \apj, 701, 1965

\bibitem[{{Neugebauer} {et~al.}(1984)}]{neugebauer84a}
{Neugebauer}, G. {et~al.} 1984, \apjl, 278, L1

\bibitem[{{Nguyen} {et~al.}(2010)}]{nguyen10a}
{Nguyen}, H.~T. {et~al.} 2010, \aap, 518, L5

\bibitem[{{Noeske} {et~al.}(2007)}]{noeske07a}
{Noeske}, K.~G. {et~al.} 2007, \apjl, 660, L43

\bibitem[{{Nordon} {et~al.}(2012)}]{nordon12a}
{Nordon}, R. {et~al.} 2012, \apj, 745, 182

\bibitem[{{Oesch} {et~al.}(2010)}]{oesch10a}
{Oesch}, P.~A. {et~al.} 2010, \apjl, 709, L16

\bibitem[{{Oesch} {et~al.}(2013)}]{oesch13a}
---. 2013, \apj, 773, 75

\bibitem[{{Oliver} {et~al.}(2012)}]{oliver12a}
{Oliver}, S.~J. {et~al.} 2012, \mnras, 424, 1614

\bibitem[{{Ouchi} {et~al.}(2004)}]{ouchi04a}
{Ouchi}, M. {et~al.} 2004, \apj, 611, 660

\bibitem[{{Overzier} {et~al.}(2011)}]{overzier11a}
{Overzier}, R.~A. {et~al.} 2011, \apjl, 726, L7

\bibitem[{{Papovich} {et~al.}(2006)}]{papovich06a}
{Papovich}, C. {et~al.} 2006, \apj, 640, 92

\bibitem[{{Peek} \& {Graves}(2010)}]{peek10a}
{Peek}, J.~E.~G. \& {Graves}, G.~J. 2010, \apj, 719, 415

\bibitem[{{Penner} {et~al.}(2012)}]{penner12a}
{Penner}, K. {et~al.} 2012, \apj, 759, 28

\bibitem[{{Pilbratt} {et~al.}(2010)}]{pilbratt10a}
{Pilbratt}, G.~L. {et~al.} 2010, \aap, 518, L1

\bibitem[{{Poglitsch} {et~al.}(2010)}]{poglitsch10a}
{Poglitsch}, A. {et~al.} 2010, \aap, 518, L2

\bibitem[{{Reddy} {et~al.}(2012)}]{reddy12a}
{Reddy}, N. {et~al.} 2012, \apj, 744, 154

\bibitem[{{Reddy} {et~al.}(2010){Reddy}, {Erb}, {Pettini}, {Steidel}, \&
  {Shapley}}]{reddy10a}
{Reddy}, N.~A., {Erb}, D.~K., {Pettini}, M., {Steidel}, C.~C., \& {Shapley},
  A.~E. 2010, \apj, 712, 1070

\bibitem[{{Reddy} \& {Steidel}(2009)}]{reddy09a}
{Reddy}, N.~A. \& {Steidel}, C.~C. 2009, \apj, 692, 778

\bibitem[{{Reddy} {et~al.}(2006){Reddy}, {Steidel}, {Erb}, {Shapley}, \&
  {Pettini}}]{reddy06a}
{Reddy}, N.~A., {Steidel}, C.~C., {Erb}, D.~K., {Shapley}, A.~E., \& {Pettini},
  M. 2006, \apj, 653, 1004

\bibitem[{{Riechers} {et~al.}(2013)}]{riechers13a}
{Riechers}, D.~A. {et~al.} 2013, \nat, 496, 329

\bibitem[{{Rodighiero} {et~al.}(2011)}]{rodighiero11a}
{Rodighiero}, G. {et~al.} 2011, \apjl, 739, L40

\bibitem[{{Roseboom} {et~al.}(2010)}]{roseboom10a}
{Roseboom}, I.~G. {et~al.} 2010, \mnras, 409, 48

\bibitem[{{Roseboom} {et~al.}(2012)}]{roseboom12a}
---. 2012, \mnras, 419, 2758

\bibitem[{{Roseboom} {et~al.}(2013)}]{roseboom13a}
---. 2013, \mnras, 436, 430

\bibitem[{{Salpeter}(1955)}]{salpeter55a}
{Salpeter}, E.~E. 1955, \apj, 121, 161

\bibitem[{{Sanders} {et~al.}(2003){Sanders}, {Mazzarella}, {Kim}, {Surace}, \&
  {Soifer}}]{sanders03a}
{Sanders}, D.~B., {Mazzarella}, J.~M., {Kim}, D.-C., {Surace}, J.~A., \&
  {Soifer}, B.~T. 2003, \aj, 126, 1607

\bibitem[{{Schlegel} {et~al.}(1998){Schlegel}, {Finkbeiner}, \&
  {Davis}}]{schlegel98a}
{Schlegel}, D.~J., {Finkbeiner}, D.~P., \& {Davis}, M. 1998, \apj, 500, 525

\bibitem[{{Scott} {et~al.}(2008)}]{scott08a}
{Scott}, K.~S. {et~al.} 2008, \mnras, 385, 2225

\bibitem[{{Scoville} {et~al.}(2007)}]{scoville07a}
{Scoville}, N. {et~al.} 2007, \apjs, 172, 1

\bibitem[{{Scoville} {et~al.}(2013)}]{scoville13a}
---. 2013, \apjs, 206, 3

\bibitem[{{Seibert} {et~al.}(2005)}]{seibert05a}
{Seibert}, M. {et~al.} 2005, \apjl, 619, L55

\bibitem[{{Shapley} {et~al.}(2003){Shapley}, {Steidel}, {Pettini}, \&
  {Adelberger}}]{shapley03a}
{Shapley}, A.~E., {Steidel}, C.~C., {Pettini}, M., \& {Adelberger}, K.~L. 2003,
  \apj, 588, 65

\bibitem[{{Smail} {et~al.}(2004){Smail}, {Chapman}, {Blain}, \&
  {Ivison}}]{smail04a}
{Smail}, I., {Chapman}, S.~C., {Blain}, A.~W., \& {Ivison}, R.~J. 2004, \apj,
  616, 71

\bibitem[{{Smail} {et~al.}(1997){Smail}, {Ivison}, \& {Blain}}]{smail97a}
{Smail}, I., {Ivison}, R.~J., \& {Blain}, A.~W. 1997, \apjl, 490, L5+

\bibitem[{{Stanway} {et~al.}(2005){Stanway}, {McMahon}, \&
  {Bunker}}]{stanway05a}
{Stanway}, E.~R., {McMahon}, R.~G., \& {Bunker}, A.~J. 2005, \mnras, 359, 1184

\bibitem[{{Stanway} {et~al.}(2007)}]{stanway07a}
{Stanway}, E.~R. {et~al.} 2007, \mnras, 376, 727

\bibitem[{{Steidel} {et~al.}(1996){Steidel}, {Giavalisco}, {Pettini},
  {Dickinson}, \& {Adelberger}}]{steidel96a}
{Steidel}, C.~C., {Giavalisco}, M., {Pettini}, M., {Dickinson}, M., \&
  {Adelberger}, K.~L. 1996, \apjl, 462, L17

\bibitem[{{Swinbank} {et~al.}(2013)}]{swinbank13a}
{Swinbank}, M. {et~al.} 2013, ArXiv e-prints

\bibitem[{{Symeonidis} {et~al.}(2013)}]{symeonidis13a}
{Symeonidis}, M. {et~al.} 2013, MNRAS, 431, 2317

\bibitem[{{Takeuchi} {et~al.}(2010){Takeuchi}, {Buat}, {Heinis}, {Giovannoli},
  {Yuan}, {Iglesias-P{\'a}ramo}, {Murata}, \& {Burgarella}}]{takeuchi10a}
{Takeuchi}, T.~T., {Buat}, V., {Heinis}, S., {Giovannoli}, E., {Yuan}, F.-T.,
  {Iglesias-P{\'a}ramo}, J., {Murata}, K.~L., \& {Burgarella}, D. 2010, \aap,
  514, A4

\bibitem[{{Takeuchi} {et~al.}(2012){Takeuchi}, {Yuan}, {Ikeyama}, {Murata}, \&
  {Inoue}}]{takeuchi12a}
{Takeuchi}, T.~T., {Yuan}, F.-T., {Ikeyama}, A., {Murata}, K.~L., \& {Inoue},
  A.~K. 2012, \apj, 755, 144

\bibitem[{{Tinsley} \& {Danly}(1980)}]{tinsley80a}
{Tinsley}, B.~M. \& {Danly}, L. 1980, \apj, 242, 435

\bibitem[{{To} {et~al.}(2014){To}, {Wang}, \& {Owen}}]{to14a}
{To}, C.-H., {Wang}, W.-H., \& {Owen}, F.~N. 2014, ArXiv e-prints

\bibitem[{{Trentham} {et~al.}(1999){Trentham}, {Kormendy}, \&
  {Sanders}}]{trentham99a}
{Trentham}, N., {Kormendy}, J., \& {Sanders}, D.~B. 1999, \aj, 117, 2152

\bibitem[{{U} {et~al.}(2012)}]{u12a}
{U}, V. {et~al.} 2012, \apjs, 203, 9

\bibitem[{{Walter} {et~al.}(2012)}]{walter12a}
{Walter}, F. {et~al.} 2012, \nat, 486, 233

\bibitem[{{Wild} {et~al.}(2011){Wild}, {Charlot}, {Brinchmann}, {Heckman},
  {Vince}, {Pacifici}, \& {Chevallard}}]{wild11a}
{Wild}, V., {Charlot}, S., {Brinchmann}, J., {Heckman}, T., {Vince}, O.,
  {Pacifici}, C., \& {Chevallard}, J. 2011, MNRAS, 417, 1760

\end{thebibliography}
\end{document}